\documentclass[review=false, screen, manuscript,nonacm]{acmart}
\AtBeginDocument{%
  }

\setcopyright{acmlicensed}
\copyrightyear{2025}
\acmYear{2025}
\acmDOI{XXXXXXX.XXXXXXX}

\usepackage{float}
\usepackage{subcaption}
\usepackage{tabularx}

\usepackage{color}

\usepackage{soul}

\usepackage{xcolor}
\usepackage{blindtext}

\usepackage{xpatch}
\makeatletter
\xpatchcmd{\ps@firstpagestyle}{Manuscript submitted to ACM}{}{\typeout{First patch succeeded}}{\typeout{first patch failed}}
\xpatchcmd{\ps@standardpagestyle}{Manuscript submitted to ACM}{}{\typeout{Second patch succeeded}}{\typeout{Second patch failed}}    \@ACM@manuscriptfalse
\makeatother
\renewcommand\footnotetextcopyrightpermission[1]{} 
\setcopyright{none}

\begin{document}

\title{Investigating the Luna--Terra collapse through the temporal multilayer graph structure of the ethereum stablecoin ecosystem}


\author{Cheick Tidiane Ba}
\email{c.ba@qmul.ac.uk}\orcid{0000-0002-4035-7464}
\affiliation{%
  \institution{Queen Mary University London}
  \city{London}
  \country{United Kingdom}
}
\affiliation{%
  \institution{University of Milan}
  \city{Milan}
  \country{Italy}
}

\author{Benjamin A. Steer}\orcid{0000-0001-9446-5690}
\email{ben.steer@pometry.com}
\affiliation{%
  \institution{Pometry Ltd}
  \city{London}
  \country{United Kingdom}
}

\author{Matteo Zignani}
\email{matteo.zignani@unimi.it}\orcid{0000-0002-4808-4106}
\affiliation{%
  \institution{University of Milan}
  \city{Milan}
  \country{Italy}
}

\author{Richard G. Clegg}\email{r.clegg@qmul.ac.uk}\orcid{0000-0001-7241-6679}
\affiliation{%
  \institution{Queen Mary University London}
  \city{London}
  \country{United Kingdom}
}

\renewcommand{\shortauthors}{Ba, et al.}

\begin{abstract}
Blockchain technology and cryptocurrencies have garnered considerable attention over the past fifteen years. The term Web3 (sometimes Web 3.0) has been coined to define a possible direction for the web based on the use of decentralisation via blockchain. Cryptocurrencies are characterised by high market volatility and susceptibility to substantial crashes, issues that require temporal analysis methodologies able to tackle the high temporal resolution, heterogeneity and scale of blockchain data. While existing research attempts to analyse crash events, fundamental questions persist regarding the optimal time scale for analysis, differentiation between long-term and short-term trends, and the identification and characterisation of shock events within these decentralised systems.

This paper addresses these issues by examining cryptocurrencies traded on the Ethereum blockchain, with a spotlight on the crash of the stablecoin TerraUSD (UST) and the currency LUNA designed to stabilise it. Utilising complex network analysis and a multi-layer temporal graph allows the study of the correlations between the layers representing the currencies and system evolution across diverse time scales. The investigation sheds light on the strong interconnections among stablecoins pre-crash and the significant post-crash transformations. We identify anomalous signals before, during, and after the collapse, emphasising their impact on graph structure metrics and user movement across layers.

This paper is novel in its use of temporal, cross-chain graph analysis to explore a cryptocurrency collapse. It emphasises the importance of temporal analysis for studies on web-derived data. In addition, the methodology shows how graph-based analysis can enhance traditional econometric results. Overall, this research carries implications beyond its field, for example for regulatory agencies aiming to safeguard users could use multi-layer temporal graphs as part of their suite of analysis tools. 

\end{abstract}



\received{10 January 2024}
\received[accepted]{28 February 2025}

\maketitle

\section{Introduction}
\label{sec:introduction}

Analysing multi-layer, temporal networks at scale is an important problem in the study of complex networks. In particular, there are issues of understanding the best time-scale to analyse, how to determine long-term and short-term trends and how to recognise and characterise shock events in such a system. In this paper, we look at six cryptocurrencies traded over the Ethereum blockchain. We use multi-layer, temporal network analysis techniques which allow us to simultaneously look at how temporal networks evolve in time and at different time-scales while considering several network layers together.

Digital currencies remain an important but potentially volatile market. In particular, there is the possibility that a particular currency will completely crash, wiping the value of that coin to a tiny fraction of its previous trade value: these events can destroy investments and savings, impacting the lives of investors. It is, therefore, important to analyse the events that lead up to a crash and to look at whether the effects of a crash can destabilise other currencies. ``Stablecoins" are an attempt to make cryptocurrencies that can be used for trading rather than speculation by ``pegging" the value of the coin to a fixed exchange rate (often one dollar to one stablecoin). The stablecoin TerraUSD (UST)\footnote{Throughout this paper we will refer to currency by the shortened name it was commonly traded under.} was pegged to trade for \$1 and was associated with a companion currency LUNA that was intended to stabilise it. Over a few days in May 2022 both UST and LUNA had collapsed and never regained their value. The collapse is believed to have been precipitated by deliberate and coordinated large-scale sales of UST in an attempt to crash UST, drive down the price of Bitcoin (BTC) and derive profit from this. Some authors~\cite{R1S1,R1S2} claim that it was the collapse of the Luna--Terra ecosystem that led to the well-known failure of FTX. 

In this paper, we use techniques from complex network analysis and, in particular, temporal multi-layer analysis to look at the events leading up to and immediately following the collapse of LUNA and UST. UST and LUNA were traded on a cryptocurrency platform known as Terra and this data is not available to us. However, the two currencies were also traded in a ``wrapped" form on the Ethereum platform, USTC (US Terra classic) and WLUNC (wrapped Luna classic) respectively. From the available set of Ethereum trades, we look at USTC and WLUNC trades and compare them with the other most highly traded stablecoins on the platform. This gives us a pool of six currencies trading over the same platform that can be compared as they evolve in the weeks before and after the collapse, hence the analysis considers 26,209,860 transactions occurring in a symmetrical analysis period around the day of the crash.

This paper uses temporal, cross-chain, graph analysis to look at the collapse of a cryptocurrency. The analysis in this paper can only be achieved with multi-layer temporal graph systems as it is necessary to look at correlations between different graph layers and to consider the evolution of the system at different time scales. We show that the six currencies are initially highly correlated in terms of the number of sales but, post-crash, separate into two groups where the ``failed" currencies correlate with each other but not the other four. We show that two high-volume selling events on UST had a significant effect on the structure of sales volumes before the crash (one event exactly coinciding with the crash) effectively decoupling its sales volume from all other currencies. We also characterise a significant ``rally" event that temporarily caused large changes to the UST and LUNA ecosystems shortly after the collapse. The collapse event has significant effects on graph metrics such as graph density and average degree but also effects on measures like clustering coefficient and weakly connected components. 


\section{Background}
\label{sec:background}

Stablecoins are digital currencies that attempt to peg their value to some other reference asset, for example the US dollar or a fixed amount of gold \cite{ante2023systematic}. This is designed to reduce the volatility of the currency as the high volatility of some currencies can put users off adopting them for use as currency (as opposed to speculation). Many different stablecoins exist traded over several different blockchain systems. Digital currencies can be traded between different blockchains by a process known as ``wrapping". In this process, a currency token originating on chain A is held in place and a wrapped coin representing the same token is minted on chain B \cite{Caldarelli2022Wrapping}. This allows, for example, BTC to be spent on the Ethereum chain as Wrapped Bitcoin (WBTC). 

\begin{table}[ht!]
    \centering
    \begin{tabular}{cccc}
     \hline\hline
         Acronym & Short description & Comment & \\
         \hline
         LUNA & Luna & Cryptocurrency, unwrapped version, not in our dataset.&\\
         UST & Terra USD & Stablecoin, unwrapped version, not in our dataset.&\\
         BTC & Bitcoin &  Cryptocurrency, not part of this study.&\\
         WLUNC & Wrapped Luna Classic & Wrapped cryptocurrency &\\
         USTC & Terra USD Classic & Wrapped stablecoin &\\
         USDT & Tether & Stablecoin &\\
         USDC & USD Coin & Stablecoin &\\
         DAI & Dai & Stablecoin &\\
         USDP & Pax Dollar & Stablecoin &\\
        \hline\hline

    \end{tabular}
    \caption{Currencies under study, the acronym is the label that the currency uses for trading.}
    \label{tab:currencies}
\end{table}

In this paper, we wish to study the ecosystem around the currencies Terra and Luna which ran on their own blockchain known as Terra. TerraUSD(UST) was a stablecoin pegged at one US dollar. Luna (LUNA) was a cryptocurrency designed to ensure the stability of the Terra stablecoin. This was achieved by an algorithmic process known as ``Anchor", that automatically would swap Terra with Luna currency to adjust their prices: in fact, the protocol kept the UST exchange rate stable by varying the pool of Terra available~\cite{briola2023anatomy}, generating or destroying tokens when needed. The Terra blockchain data is unavailable to us, however, we can study wrapped versions of UST and LUNA traded on the Ethereum blockchain. We provided a summary of the currencies under study in Table \ref{tab:currencies}.

The interest in studying the Terra/Luna ecosystem comes from the sudden complete crash of UST in May 2022. A timeline extracted from \cite{briola2023anatomy} and \cite{shamsi2022chartalist} is as follows:
\begin{itemize}
    \item 3rd April 2022 \textbf{(S1)}: Anomalous UST selling event discovered during this analysis. We denote this event as sale 1 (S1).
    \item 19th April 2022 \textbf{(S2)}: Anomalous UST selling event discovered during this analysis. We denote this event as sale 2 (S2).
    \item 20th April 2022: UST becomes the third largest stablecoin. 
    \item 5th May 2022: The start of strong and persistent selling pressure is evident on BTC and LUNA.
    \item 7th May 2022: Agreement is made to buy a large number of UST for BTC greatly reducing the liquidity in the pool. Simultaneously a large number of UST tokens are put up for sale (which caused the first UST de-pegging, below \$0.99 \cite{briola2023anatomy}).
    \item 8th May 2022: UST loses its peg against the dollar falling to \$0.99 and the Luna Foundation group uses reserve funds in BTC to prop up the currency.	
    \item 9th May 2022 \textbf{(C)}: UST depegs again falling in value to \$0.35 leading to customers trying to sell their reserves to exit the market. We denote this event as a crash (C). 
    \item 10th May 2022: Luna Foundation Guard sells reserve Bitcoin (BTC) to attempt to restore the peg.
    \item 12th May 2022: LUNA has collapsed by almost 99\%.	
    \item 13th May 2022: Terra temporarily halts its Blockchain and Binance suspends LUNA trading. 
    \item 25th May 2022: Proposal \cite{luna_proposal_2022,news_proposal_1623_terra_2022,medium_mc_terra_2022} for the launch of a new blockchain Terra 2.0 is official.  
    \item 27th May 2022 \textbf{(T2)}: Terra 2.0 is live. Original LUNA tokens became Luna Classic (LUNC). The original UST stablecoin became TerraClassicUSD (USTC). Unlike the original Terra network, Terra 2.0 has no algorithmic stablecoin \cite{TerraNewBlockchain}. We denote this event as Terra 2.0 (T2).
\end{itemize}

The final consequence was not only the collapse of the Terra platform, UST and LUNA but also a decrease in the price of BTC. The motivation speculated for the attack is that the attackers were short-selling large amounts of BTC.
Short selling is, in effect, a bet that a traded asset will decrease in price. The short seller contracts to borrow and sell the asset now and buy it at a later date at market price. If the price drops the seller, has profited by the amount of the price drop. In some markets, short selling is highly regulated as it has been implicated in market crashes~\cite{jiang2022short}. In our case study, there is a belief that high levels of selling of Bitcoin observed in the days prior to the crisis were linked to short selling~ \cite{briola2023anatomy}.


\paragraph{Related works}

The Luna crash serves as an interesting case study due to its relevance within the stablecoins ecosystem, attracting attention from both industry and the research community.
There are many works describing the ecosystem and the properties of the Terra system. For instance, Uhlig \cite{uhlig2022luna} proposes a theoretical framework for the analysis of hourly data on prices and market capitalisation data, delineating the gradual unfolding of the crash. Limited research leverages on-chain data. Cho \cite{cho2023token} investigates the underlying causes of the de-pegging event focusing on the token economics of the Terra Blockchain. They utilise on-chain data focusing on transactions converting USTC to LUNC over the four days following the May 9 crash, identifying critical design flaws in the conversion process exploited by attackers. These works focus mainly on Terra. However, various studies explore the impact of the crash on other currencies. 
De Blasis \textit{et al.}~\cite{de2023intelligent} examine the impact of the USTC collapse on other stablecoins by using proprietary minute-by-minute price transaction data of various cryptocurrencies, over a symmetrical 40-day period around the USTC crash of May 9th. Their work shows the presence of significant contagion from the USTC collapse, which could be explained by traders' herding behaviour i.e. users reacting to the crash by moving to other currencies. Briola \textit{et al.}~\cite{briola2023anatomy} provide a background on the Terra project and describe the main events that led to the crash, before delving into an analysis focused on the hourly price data time series. They study the evolution of dependencies between 61 cryptocurrencies, by leveraging the price time series to define a network structure among currencies, a correlation network: their analysis seems to exclude the presence of herding behaviour. The importance of analyzing cross-chain behaviour is denoted by other works as well. Barthere \textit{et al.}~\cite{barthere2022chain} analyses proprietary data from Terra and Ethereum to identify the main users involved in the crash. They assert that a small subset of users initiated the crash by leveraging system vulnerabilities, benefiting significantly by converting USTC to other coins before the crash. Liu \textit{et al.}~\cite{liu2023anatomy} employ a similar dataset and approach, leveraging proprietary transaction data from Terra and Ethereum. Their extensive analysis across multiple chains and assets reveals the complex nature of the run on Terra. At the same time, they believe the crash was exacerbated by concerns about system sustainability and not just a concentrated market manipulation. Although they have user transaction data, in their study, the authors do not focus on the transaction networks but focus more on describing the events and identifying the main actors. Meanwhile Vidal-Tom\'{a}s \textit{et al.}~\cite{R1S1} look at the wider context including the well-known FTX crash from the perspective of the Luna--Terra crash, citing the latter as a significant contributing factor to the former. Conlon \textit{et al.}~\cite{R1S2} take a similar view. 

A key limitation in these works is while they provide important results on the crash, they neglect the analysis of the transactions among users from a network standpoint: indeed, the circulation of tokens within the ecosystem can be effectively represented as networks or graphs~\cite{barabasi2013network} a useful tutorial is found in~\cite{TutorialKDD}. In this representation, nodes (or vertices) symbolise users, while links (or edges) denote the dynamic interactions or relationships among them. This modelling and analysis approach for currency systems through networks has significantly enhanced our understanding of financial transaction networks~\cite{aridhi2016big,coscia2021atlas}. Moreover, temporal networks \cite{holme2019temporal} have proven to be a valuable framework for comprehending web-based currency systems such as Bitcoin~\cite{meiklejohn2013fistful,wang2021large,MaesaIJDSA2018,GensollenANS2020} and Ethereum~\cite{guo2019graph,Sharma2023ACMTWEB,nadini2021mapping,galdeman2022disentangling}, as well as in the context of new-generation social media platforms~\cite{guidi2021graph,BaWebSci22,ba2023characterizing}. However, none of these works analyses the stablecoin ecosystem from a network standpoint. Recent works analyzing the Chartalist dataset by Shamsi \textit{et al.}~\cite{shamsi2022chartalist} highlighted the importance of network structure in the stablecoin system.  For instance, Huang \textit{et al.}~\cite{huang2023fast} focus on change detection on dynamic graphs, proposing a spectral method to create dynamic graph embeddings and an associated anomaly score can be used to detect significant changes. The main result is that they detected anomalies in the graph embeddings three days before the crash; however, the work used a unique graph, without distinguishing the stablecoins, hence the detected anomalies before it may not necessarily be related to the Luna Terra system or the crash.
Similarly, Zhu \textit{et al.}~\cite{zhu2023core}, focuses on the task of trend detection in blockchain-based networks, with a method focused on core decomposition to detect a change in the activity in the temporal transaction network. Among the selected datasets for evaluation, they rely on a single dataset: the authors can detect a perturbation in network structure caused by the crash, but they too model the entire transaction data as a homogeneous temporal network. 

While network structure seems an important factor to analyse, none of the current works analyse the transaction network in relation to the crisis; and those who leverage network structure failed to properly separate the different currencies, treating the transactions across different currencies as the same object of the study.

\section{Research motivation}
\label{sec:rqs}

The cryptocurrency and stablecoin ecosystem here consists of several heavily interlinked currencies that can be studied together both before and after the shock. This paper studies the effects of the shock on the currencies most directly affected but also on those that are related. This is consistent with the suspicion that the attack on Terra/Luna was intended as an attack on the price of BTC. The paper answers the following research questions:

\paragraph{What is the nature of the cryptocurrency ecosystem before and after the attack} Can we see commonalities in the signals between the different currencies? Are users investing in several currencies or specialising in one? We analyse the cross-correlation of various measures (eg number of trades and number of tokes traded) between currencies to ask whether similar patterns of selling and buying were occurring over time within the system. 

\paragraph{Can we anticipate the crash by looking at trade volumes of cryptocurrencies in the days leading up to it? What are the effects of the crash on trade volume?} Are the events preceding it unusual in scale or characteristics? Are they isolated to just one currency? We look at the volume and number of trades in each currency over the immediate time period before the attack to find those unusual ``whale" events that precede the crash and the aftershocks from the event. 

\paragraph{What are the effects of the attack on the system on the most directly affected currencies?} We look at the size and number of trades after the crash event to consider the effects of the crash focusing on the most affected currencies and considering different time windows. 

\paragraph{Can the after effects of the attack be seen on graph structure as a whole?} We look at traditional graph measures such as reciprocity, clustering coefficients, weakly connected components, mean degree and graph density on each currency treated as an individual graph. We ask whether there are short and longer term trends visible in these graph measures as a result of the crash.

\paragraph{Where did user interest move post-crash?} We look at how users changed their investment strategies after the crash event. Did users of currencies not directly affected change their behaviour? Where did users in the most directly affected currencies move their investments?

\paragraph{Can we locate individuals associated with the crash?} We will deepen the analysis of user activity on days characterised by anomalous user activity. Are the anomalies detected caused by a few individuals or the result of a widespread change of behaviour? Are there common actors active during the anomalous events?

\section{Methodology}
\label{sec:method}
In this section we give a description of our dataset and the methodology we will use for analysis. It should be remembered throughout this and the next section that the analysis of USTC and WLUNC are looking only at those transactions that can be seen on the Ethereum blockchain. The number of transactions on the Terra blockchain (that we cannot observe) where Terra and Luna are traded natively are likely to be far larger in volume. Our analysis is conducted using the open-source, temporal graph software Raphtory developed at Queen Mary University of London and Pometry Ltd~\cite{steer2023raphtory}. The software allows us to conduct analysis across multiple layers, over multiple time windows simultaneously. 

\subsection{Dataset}
\label{sec:dataset}

The dataset we rely on is the stablecoins dataset presented in Chartalist datasets by Shamsi \textit{et al.}~\cite{shamsi2022chartalist}. It covers the activity of users of the top stablecoins traded on Ethereum ordered by market capitalisation at the time of the crash, alongside USTC and WLUNC. The dataset consists of 70,678,297 transactions, occurring between 1st April 2022 and 1st October 2022, covering a month prior to the crash and six months after the crash. We are focused on specific details, namely the sender ( from\_address) and the receiver (to\_address) of the currency, the transfer timestamp (time\_stamp), the token amount sent (value), and the smart contract address (contract\_address) that identifies the currency traded. The currencies under examination are associated with the Terra Ecosystem: USTC and WLUNC\footnote{In the dataset, WLUNC is denoted as LUNC, but the contract address points to Wrapped LUNA https://etherscan.io/token/0xd2877702675e6ceb975b4a1dff9fb7baf4c91ea9, currently traded as WLUNC }. Additionally, the dataset encompasses user transactions involving other stablecoins, including Tether (USDT), a collateral-backed cryptocurrency stablecoin \cite{tether}; Dai, a USD-pegged stablecoin maintained through smart contracts, governed by MakerDAO \cite{team2020maker}; Pax Dollar (USDP), a stablecoin collateralized 1:1 by USD in Paxos-owned US bank accounts and approved by Wall Street regulators \cite{usdp_2021}; and USD Coin (USDC), managed by the Centre consortium, backed by a dollars in reserve or other undisclosed "approved investments" \cite{usd_coin_may_2018}. We provided a summary of the currencies in Table \ref{tab:currencies}.

\begin{figure}[ht] 
    \begin{minipage}[t]{0.48\textwidth}
    \centerline{\includegraphics[width=0.99\textwidth]{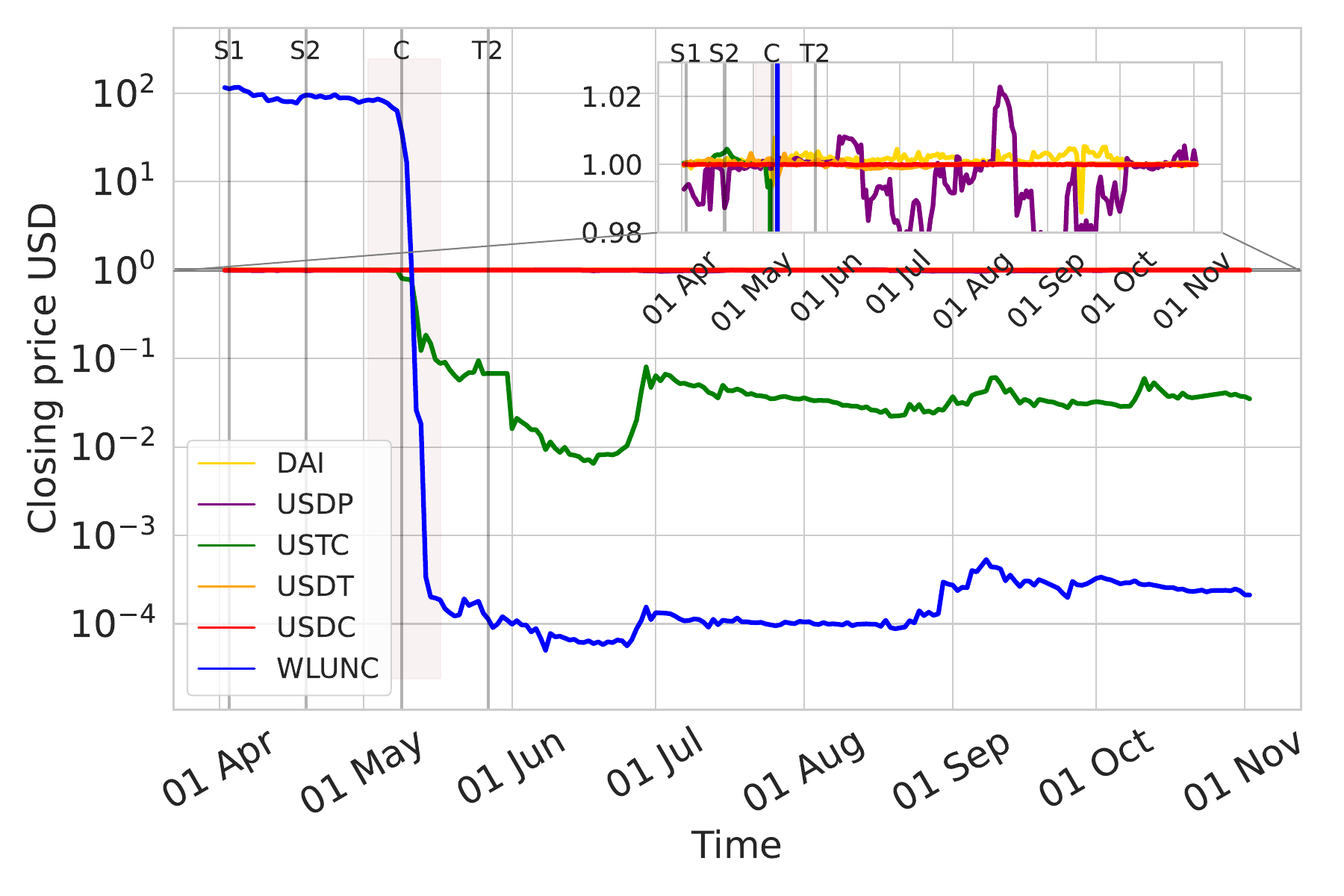}}
    \subcaption{Close values log}
    \label{fig:Close-values-log}
    \end{minipage}
    \begin{minipage}[t]{0.48\textwidth}
    \centerline{\includegraphics[width=0.99\textwidth]{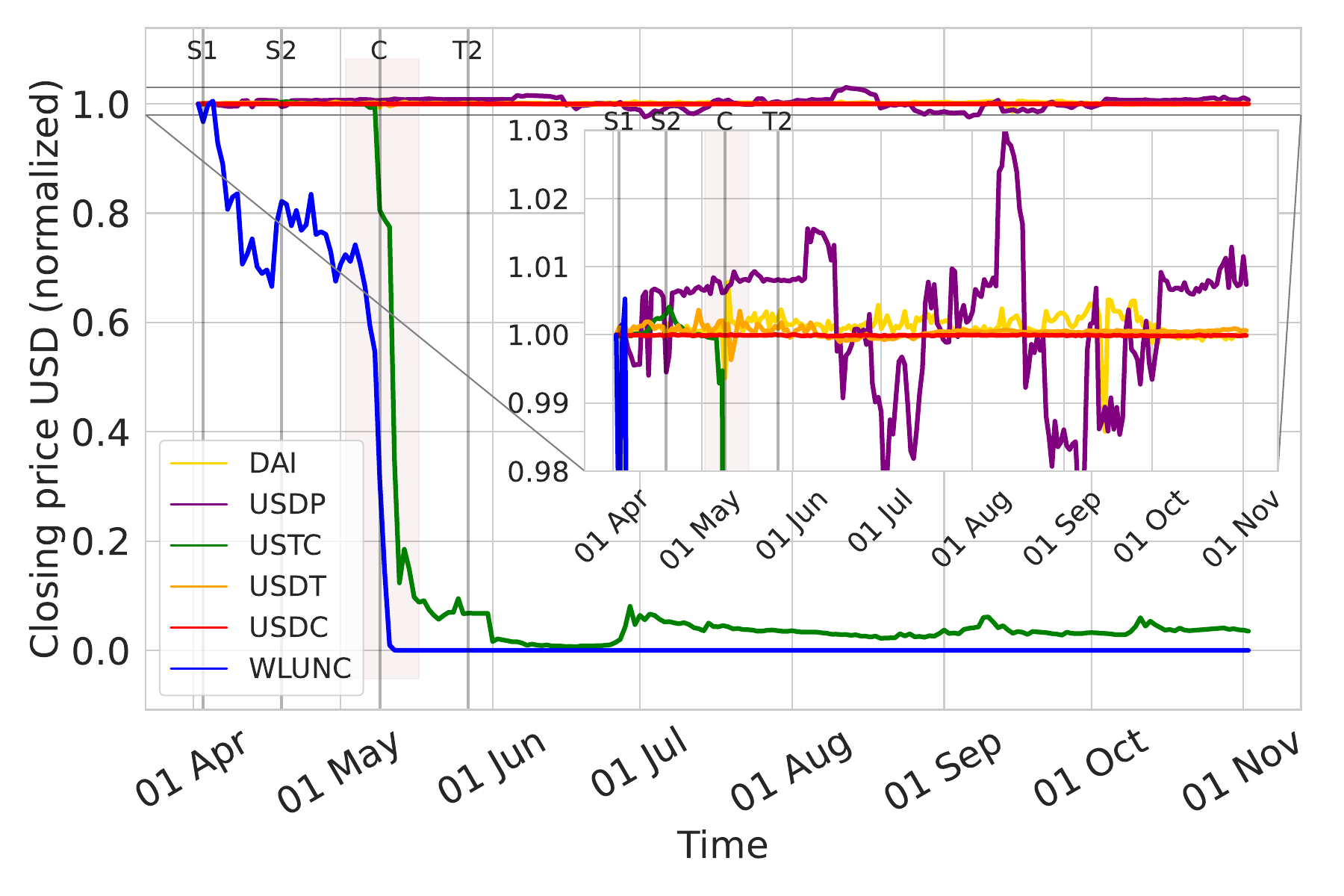}}
    \subcaption{Close values norm starting value}
    \label{fig:Close-values-norm-starting-value}
    \end{minipage}
\caption{Daily price of the currencies under study, in the periods covered by the full dataset. We represent the closing prices on April 1st and onward, normalising the values of each by dividing by the starting value in the observed period. In the insets, we provide a zoomed version around the value of 1 USD.}
\label{fig:closing}
\end{figure}

We also integrate the available data with currency daily prices. We recover the price data from the web platform CoinCodex \cite{coincodex}. The obtained data is presented in \figurename~\ref{fig:closing}. From \figurename~\ref{fig:Close-values-log} we can see how Luna was extremely valuable with a peak value surpassing \$120 a token. \figurename~\ref{fig:Close-values-norm-starting-value} presents an alternative version of the plot that gives us a better view of the stablecoins. The plot is normalized by the starting value observed in the period, highlighting the fall of USTC while other stablecoins largely exhibit their designed behaviour, holding a value very close to one dollar.

\subsection{Analysis methods}

\begin{figure}[ht] 
    \centerline{\includegraphics[width=0.4\textwidth]{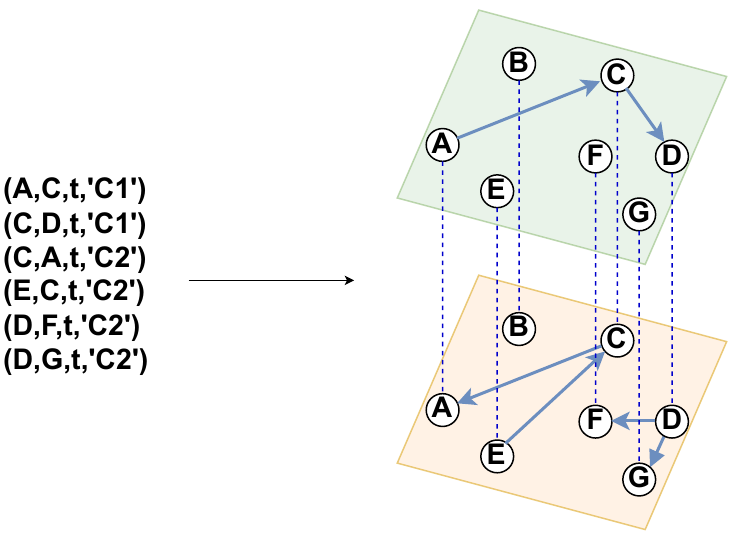}}
\caption{Illustration of the construction of the temporal multilayer network. Given a set of transactions, we can model them as a network, where each layer represents one of the currencies under study.} 
\label{fig:modeling}
\end{figure}

The data is analysed as a directed, multilayer temporal graph. A node represents an Ethereum wallet. A directed edge represents a transaction between two wallets, the weight represents the amount of currency, the direction represents the direction in which money is transferred and the layer represents the individual currency. 
This is illustrated in Figure \ref{fig:modeling}.

Using the Raphtory system \cite{steer2023raphtory} we can easily perform analysis using sliding windows to consider the trades within different parts of the data and investigate how the temporal graph behaves within a time period\footnote{Code will be made publicly available in case of acceptance.}. If we define a time window $T=\left[t_1,t_2\right)$ then that window will include trades made at all times $t \leq t_1$ and $t < t_2$.\footnote{The closed/open interval means we do not double count trades that occur exactly on a border if, for example, we move a window forward every hour we do not want trades occurring exactly on the hour to be in two different windows.} The degree of a node within a time window is a measure of the number of trading partners that the wallet has within that time window. The balance of the weights of inbound minus outbound link weights is the node's income during the period.

\begin{figure}[ht] 
    \centerline{\includegraphics[width=0.8\textwidth]{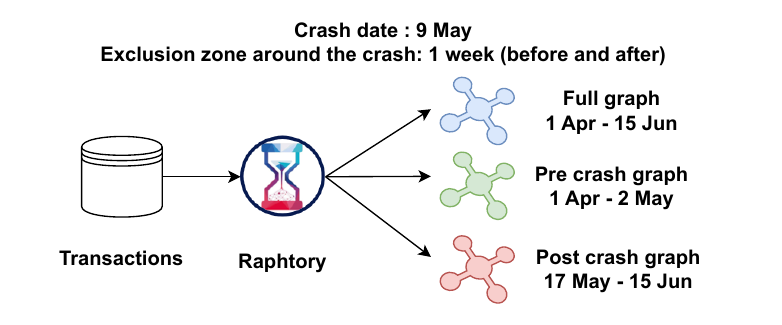}}
\caption{Methodology pipeline. The illustration summarises the selected time filters and the different temporal versions of the graphs we analyse.} 
\label{fig:methodology}
\end{figure}

For analysis purposes, we want to separate the data into a pre-crash and post-crash period to investigate the normal behaviour of the market and to see if it returns to the same situation after the crash occurs. Obviously, when such an event ends is somewhat arbitrary. For the purposes of this paper, we have chosen the 9th May 2022 as the centre of the crash event and we exclude that day and a week on either side. This leads to a fifteen-day exclusion period with the pre-crash period including all data up to the end of 1st May and the post-crash period including all data from the beginning of the 17th May. This gives us thirty days of complete data before the exclusion zone and by considering all data up to the end of 15th June we have thirty days after the exclusion zone. In Figure \ref{fig:methodology}, we present a summary of the methodology and selected filters for reproducibility.

The exclusion zone is plotted on all graphs as a light red area with $C$ marking the centre. Graphs are also marked with $S1$ and $S2$ for the two major selling events on 3rd April 2022 and 19th April 2022 that preceded the crash (see the timeline in Section~\ref{sec:background}) and with $T2$ for the launch of Terra 2.0 blockchain on 27th May 2022.



\subsection{Basic network statistics}

\begin{table}[ht]
\centering
\caption{Network statistics in the full graph and each layer. Active: source (out) or target (in) of at least one transaction.                 Sinks perform no out action, while sources perform only out actions}
\label{tab:table-stats-subgraph-active}
\begin{tabular}{cccccccc}
 \hline\hline
 & Full graph & USDC & DAI & USDT & USTC & WLUNC & USDP \\
 \hline
$|V|$ &    2,908,261 &    1,025,672 &     147,966 &    1,824,419 &      39,365 &      67,756 &       6,163 \\
$|E|$ &    5,963,549 &    2,169,372 &     288,935 &    3,451,307 &      76,016 &     106,464 &       9,115 \\
Trades total  &  26,209,860 &    10,222,791 &     1,569,005 &   13,264,318 &    394,319 &      699,345 &     60,082 \\
Active out &    2,336,953 &     870,624 &     121,329 &    1,429,503 &      32,433 &      32,906 &       5,013 \\
Active in &    2,547,837 &     907,951 &     126,998 &    1,580,164 &      37,074 &      66,339 &       4,289 \\
Sources &     360,424 &     117,721 &      20,968 &     244,255 &       2,291 &       1,417 &       1,874 \\
Sinks &     571,308 &     155,048 &      26,637 &     394,916 &       6,932 &      34,850 &       1,150 \\
 \hline
\end{tabular}
\end{table}

Considering the different currency transactions as layers in a multi-layer graph we first want to establish the size of each layer/currency in terms of nodes ($|V|$), edges ($|E|$), and total trades/transactions. This can be seen in Table~\ref{tab:table-stats-subgraph-active}. The layers exhibit considerable heterogeneity in size. It can be easily seen that USDT is by far the largest and USDP the smallest.  Notably, the two directly involved layers, USTC and WLUNC, are not among the largest layers, but they still surpass one of the layers (USDP). This observation is intriguing, especially considering that USTC and WLUNC are the wrapped versions of the original currencies. Across all layers, a predominant pattern emerges where most users are typically active either as senders or receivers, with a higher frequency of activity as receivers. Additionally, a notable portion of users exclusively serves as sources, solely sending out currency, or act as sinks, exclusively receiving currency.

\begin{figure}[ht] 
    \begin{minipage}[t]{0.325\textwidth}
    \centerline{\includegraphics[width=0.99\textwidth]{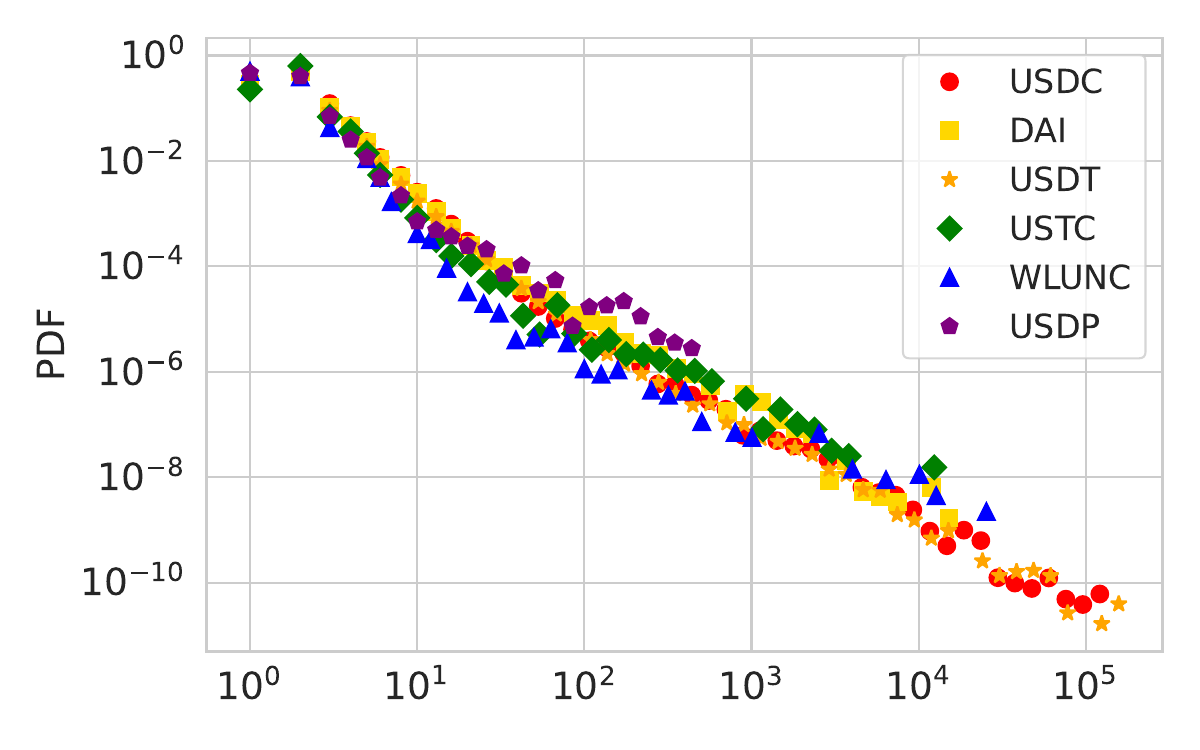}}
    \subcaption{Degree full period}
    \label{fig:Degree-full}
    \end{minipage}
    \begin{minipage}[t]{0.325\textwidth}
    \centerline{\includegraphics[width=0.99\textwidth]{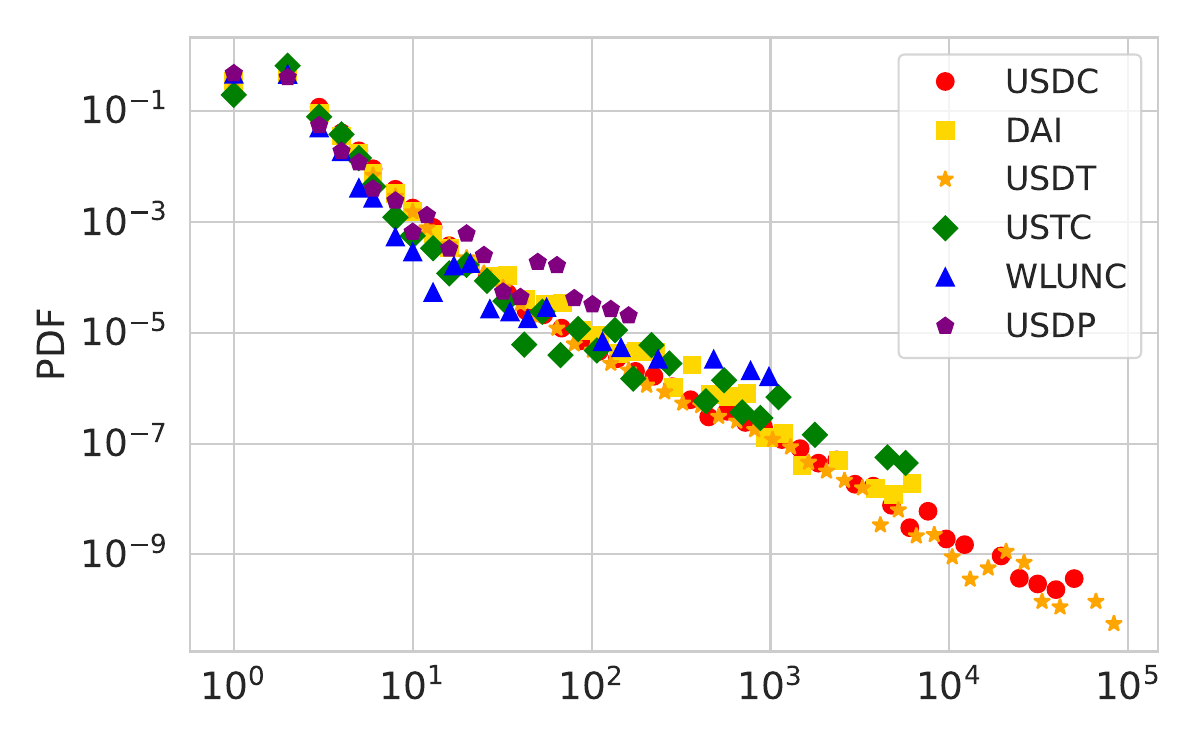}}
    \subcaption{Degree pre crash}
    \label{fig:Degree-pre}
    \end{minipage}
    \begin{minipage}[t]{0.325\textwidth}
    \centerline{\includegraphics[width=0.99\textwidth]{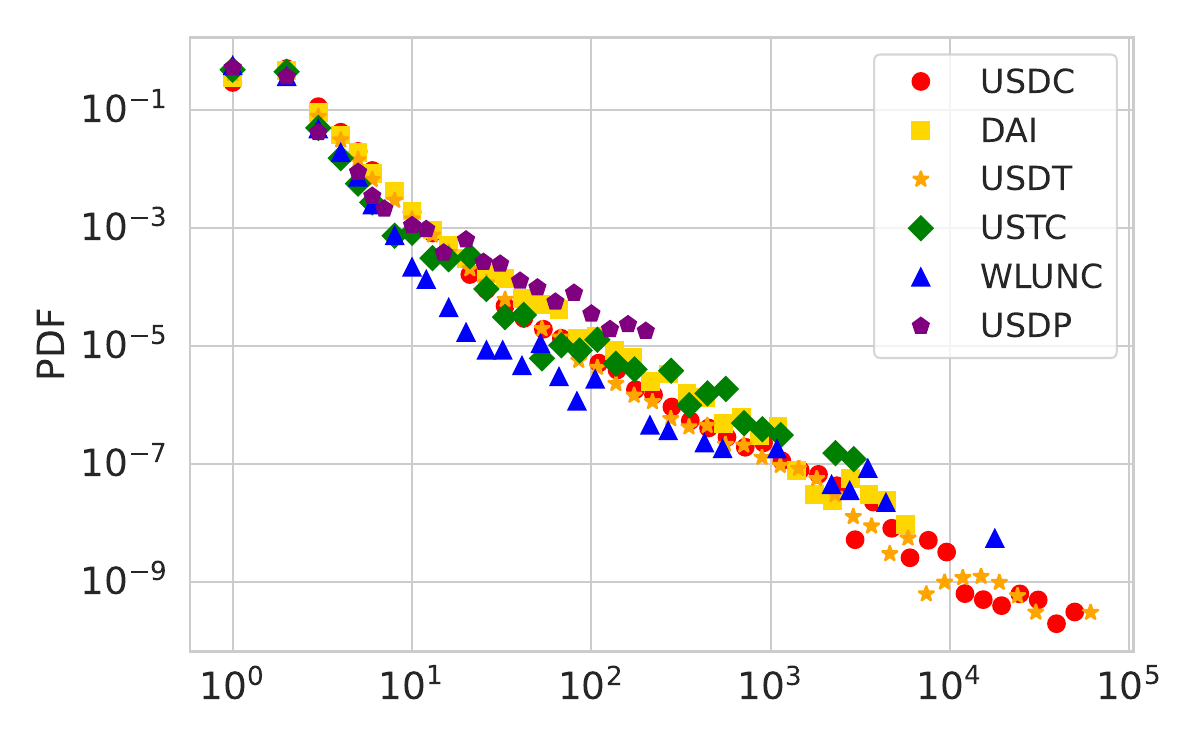}}
    \subcaption{Degree post crash}
    \label{fig:Degree-post}
    \end{minipage}
\caption{Probability distribution for node degree on the full data set, pre-crash period and post-crash period. The axes are logarithmic on both scales.}
\label{fig:degsDegree}
\end{figure}

\figurename~\ref{fig:degsDegree} shows the degree distribution of the different stable coins and WLUNC pre and post crash as well as the full data set. The degree of a node is the number of different trading partners that the node has. Currencies with a larger userbase will tend to have higher degree nodes and hence stretch to the bottom right of the probability distribution function (PDF). The full data is a larger set of trades hence we expect slightly larger node degrees to result. When comparing the pre and post crash data we can see that post crash WLUNC has a stretched distribution indicating traders with a larger number of trading partners in this period. By contrast, USTC has fewer trades in the post crash period. The other coins have quite a similar distribution pre and post crash. 

\section{Results}
\label{sec:results}
Using the methodology from the previous section we set out to answer various research questions concerning the stablecoin ecosystem.

\subsection{What is the nature of the cryptocurrency ecosystem before and after the attack}\label{results:rq1}

Table~\ref{tab:table-layers} shows the number of users active classed by the number of layers they are active in before and after the crash and in the full data. It can be seen that most users trade only a single currency. The number of users trading in all six currencies is low. One possible reason for this is that stablecoins if they work correctly are supposed to hold a pegged value. A user who believes this will happen might have little incentive to trade multiple stablecoins. 

\begin{table}[ht]
\caption{Active users.  We represent in how many layers are users active.}
\label{tab:table-layers}
\begin{tabular}{ccccccc}
 \hline\hline
Layers & Full data users &  & Pre-crash users &  & Post-crash users & \\
 \hline
1 & 2,727,485 &      93.78 $\%$& 1,201,638 &      95.43 $\%$ & 1,305,639 &      94.86 $\%$\\
2 & 161,004 &       5.54 $\%$ & 51,758 &       4.11  $\%$& 64,523 &       4.69  $\%$\\
3 & 17,610 &       0.61  $\%$& 5,174 &       0.41  $\%$& 5,737 &       0.42  $\%$\\
4 & 1,848 &       0.06  $\%$& 470 &       0.04  $\%$& 362 &       0.03  $\%$\\
5 & 258 &       0.01  $\%$& 65 &       0.01  $\%$& 87 &       0.01  $\%$\\
6 & 56 &       0.00  $\%$& 26 &       0.00  $\%$& 29 &       0.00  $\%$\\
 \hline
\end{tabular}
\end{table}

\begin{figure}[ht!] 
    \begin{minipage}[t]{0.99\textwidth}
\centerline{\includegraphics[width=0.7\textwidth]{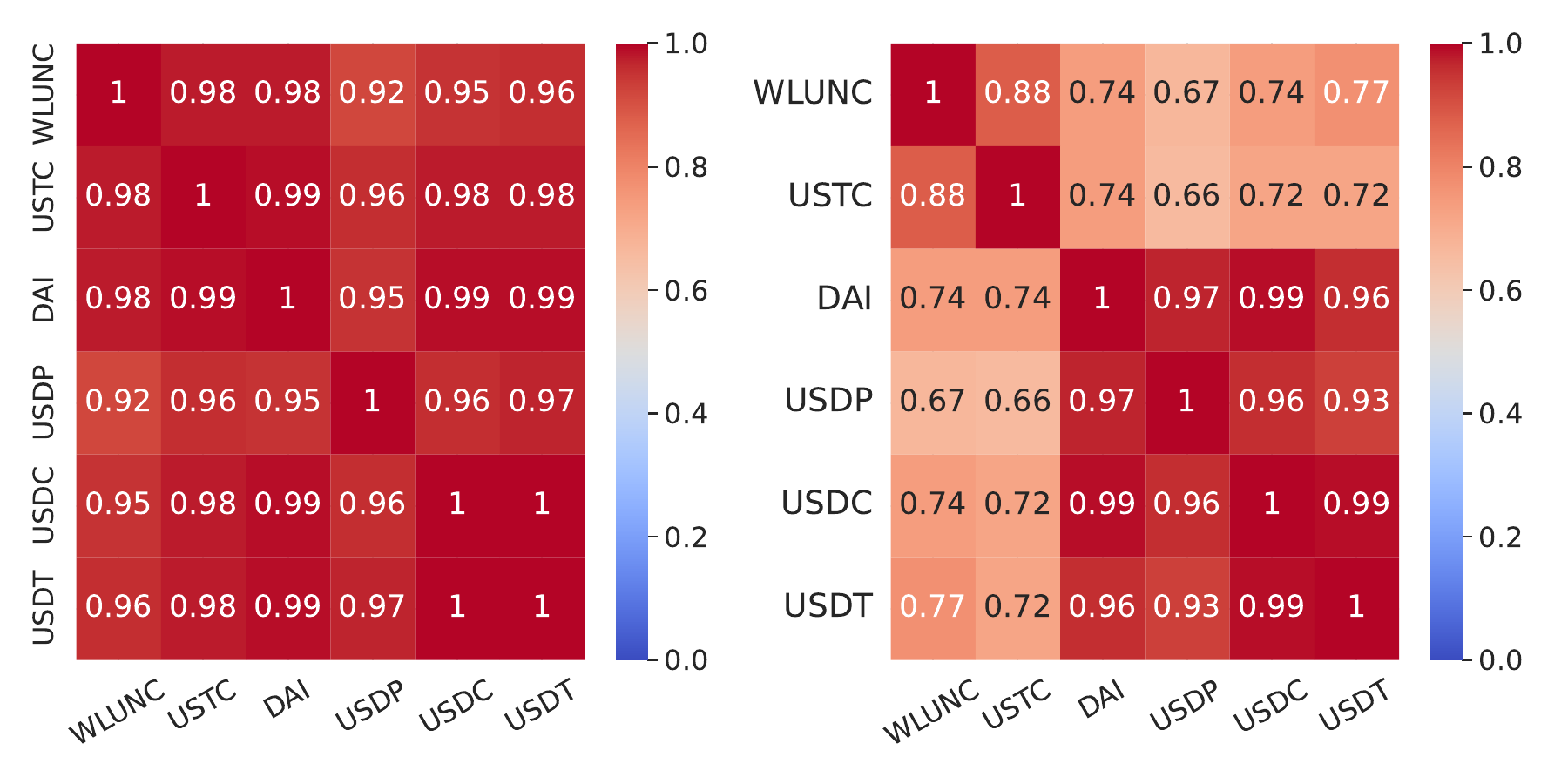}}
\subcaption{Cross-correlation $\max_k(\rho(k))$ for the raw number of transactions per day pre-crash (left), post-crash (right).}
\label{fig:heatmap-corrs-Transactions count}
\end{minipage}
    \begin{minipage}[t]{0.99\textwidth}
\centerline{\includegraphics[width=0.7\textwidth]{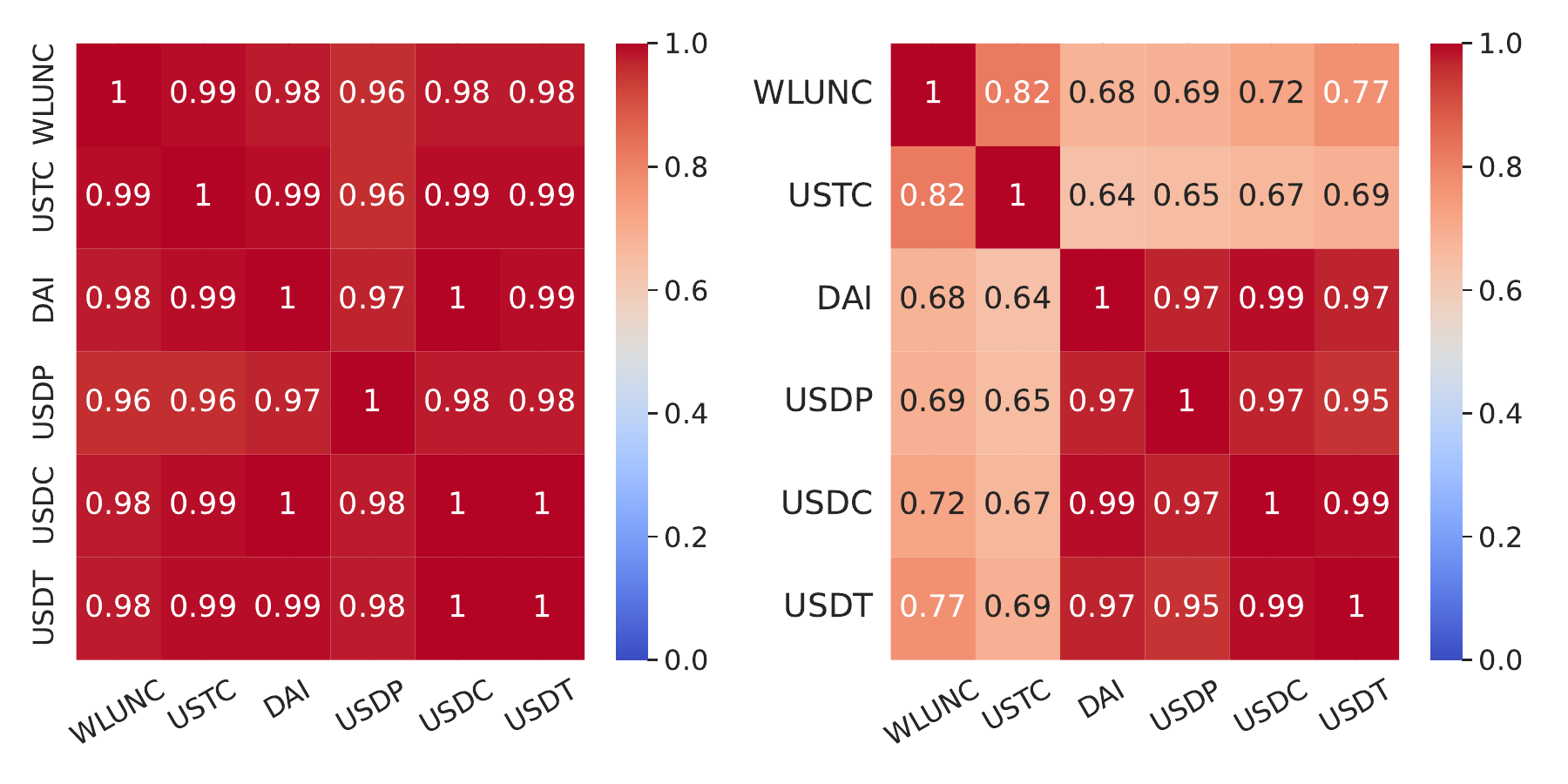}}
\subcaption{Cross-correlation $\max_k(\rho(k))$ for the number of graph edges per day pre-crash (left), post-crash (right).}
\label{fig:heatmap-corrs-Edges}
\end{minipage}
\begin{minipage}[t]{0.99\textwidth}
\centerline{\includegraphics[width=0.7\textwidth]{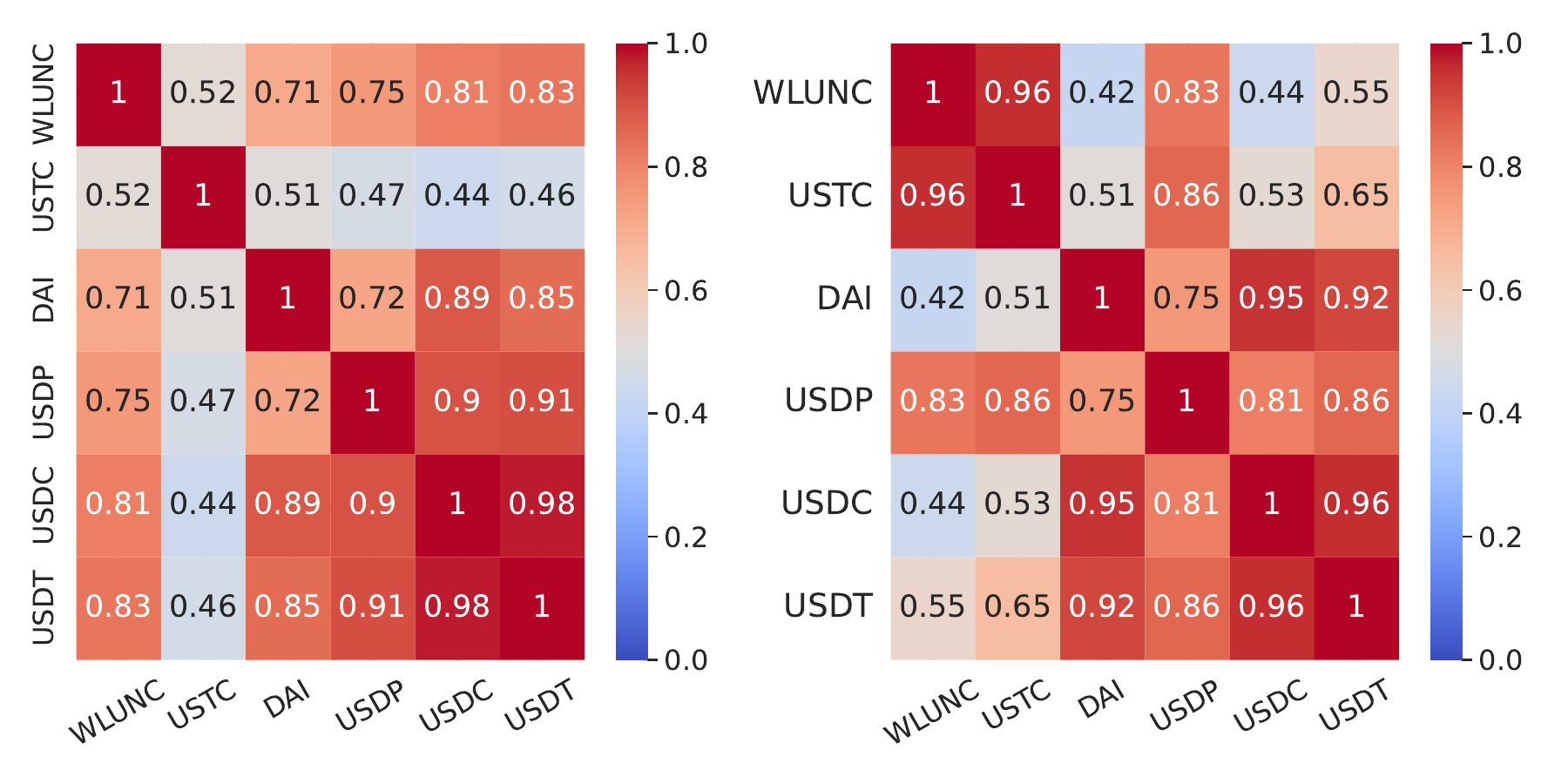}}
\subcaption{Cross-correlation $\max_k(\rho(k))$  for the total trade value per day pre-crash (left), post-crash (right).}
\label{fig:heatmap-corrs-Selling USD}
\end{minipage}
\caption{Cross-correlation $\max_k(\rho(k))$  of trade patterns for the six currencies studied.}
\label{fig:heatmap-corrs}
\end{figure}

An obvious question to ask about the system is whether the trading patterns within one currency correlate with those patterns within another currency. To do this for every day we extract the raw number of edges in the graph, the raw number of transactions and the total value of transactions converted to USD. This gives us some different time series. We use the normalised cross-correlation function $\rho(k) \in (-1,1)$ where $k$ is the lag in days. Here $\rho(k)=1$ is perfect correlation, $\rho(k)=-1$ a perfect anti-correlation and $\rho(k)=0$ no correlation at all. We plot the maximum $\max_k(\rho(k))$ for all $k$ as a matrix of heatmaps in \figurename~\ref{fig:heatmap-corrs}, with separate heatmaps for the pre-crash and post-crash. In almost all cases this occurred when $k=0$. 

From the analysis of \figurename~\ref{fig:heatmap-corrs} the most obvious conclusion is that many of the currencies have an extremely high correlation in the pre-crash period. The number of trades and graph edges (pairs of trading partners) show extremely high correlation in this period. Some degree of correlation might be expected since currency fluctuations are driven by the same external events (for example earlier crashes). However, the extremely high correlations in the left hand side of \figurename~\ref{fig:heatmap-corrs}(a) and (b) are remarkably high and this correlation can be clearly seen looking at the raw transaction traces on the left hand side of \figurename~\ref{fig:str-1day-rolling}(a). The graph of total trade value shows less correlation even in the pre-crash period but the currency USDT is a marked anomaly having low correlation compared with the others. The likely reason for this is a small number of large volume trade events in the pre-crash period, that appears to be part of the attempt to destabilise the WLUNC, USDT system. In the post crash period, the system has split into two with the remaining four stablecoins extremely highly correlated with each other, but USDT and WLUNC have a much lower correlation. It is also interesting to note that WLUNC and USDT correlate well with each other in almost all analyses apart from the pre-crash trade value which is disrupted for USDT as previously described. The major conclusion of this analysis is that pre-crash all currencies exhibited extremely similar trade patterns (by this measure at least) but post crash WLUNC and USDT have become a separate eco-system. 

\subsection{Can we anticipate the crash by looking at the cryptocurrencies in the days leading up to it?}

\begin{figure}[ht!] 
    \begin{minipage}[t]{0.48\textwidth}
    \centerline{\includegraphics[width=0.99\textwidth]{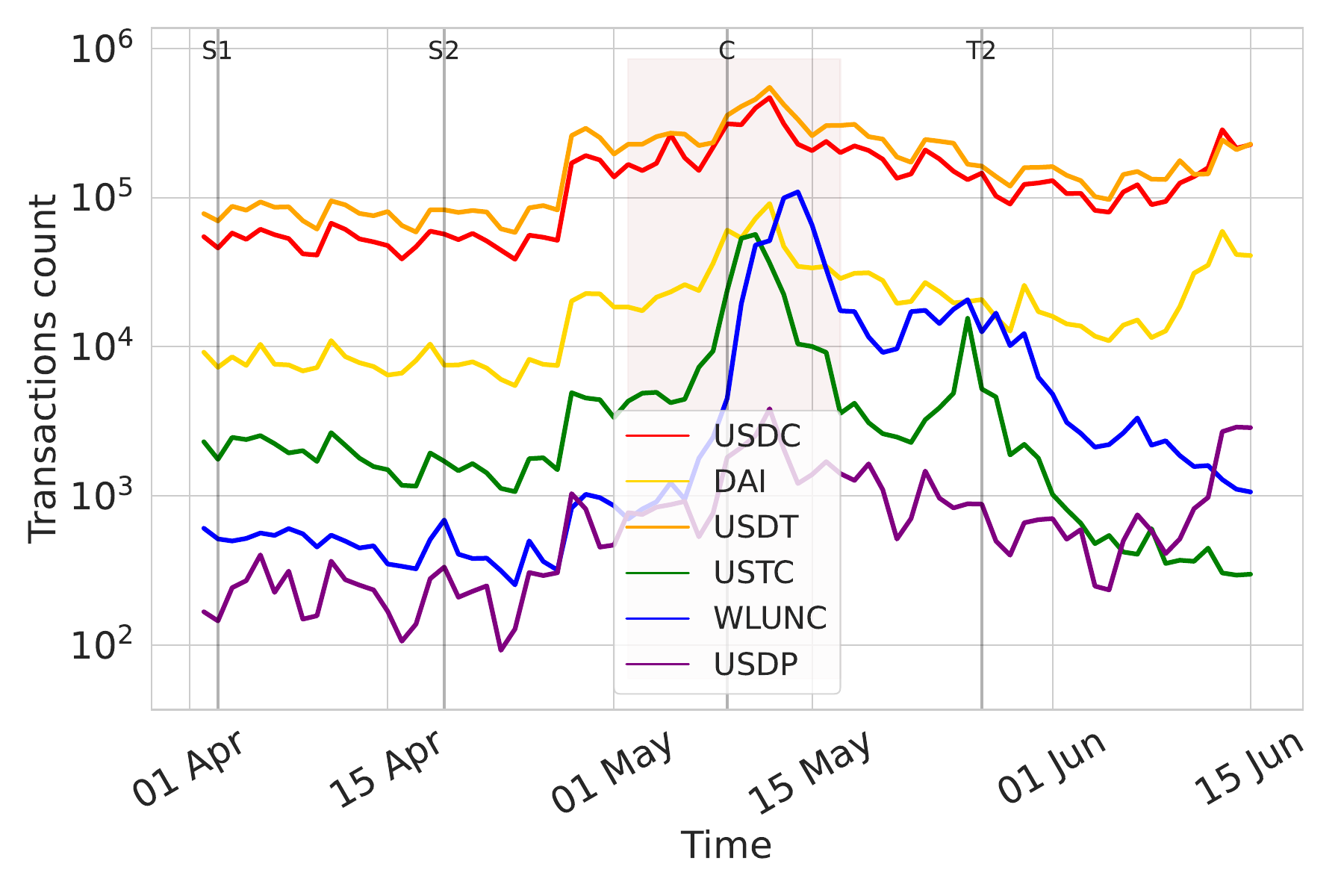}}
    \subcaption{Transactions each day}
    \label{fig:Tnx-Count-1 day rolling}
    \end{minipage}
    \begin{minipage}[t]{0.48\textwidth}
    \centerline{\includegraphics[width=0.99\textwidth]{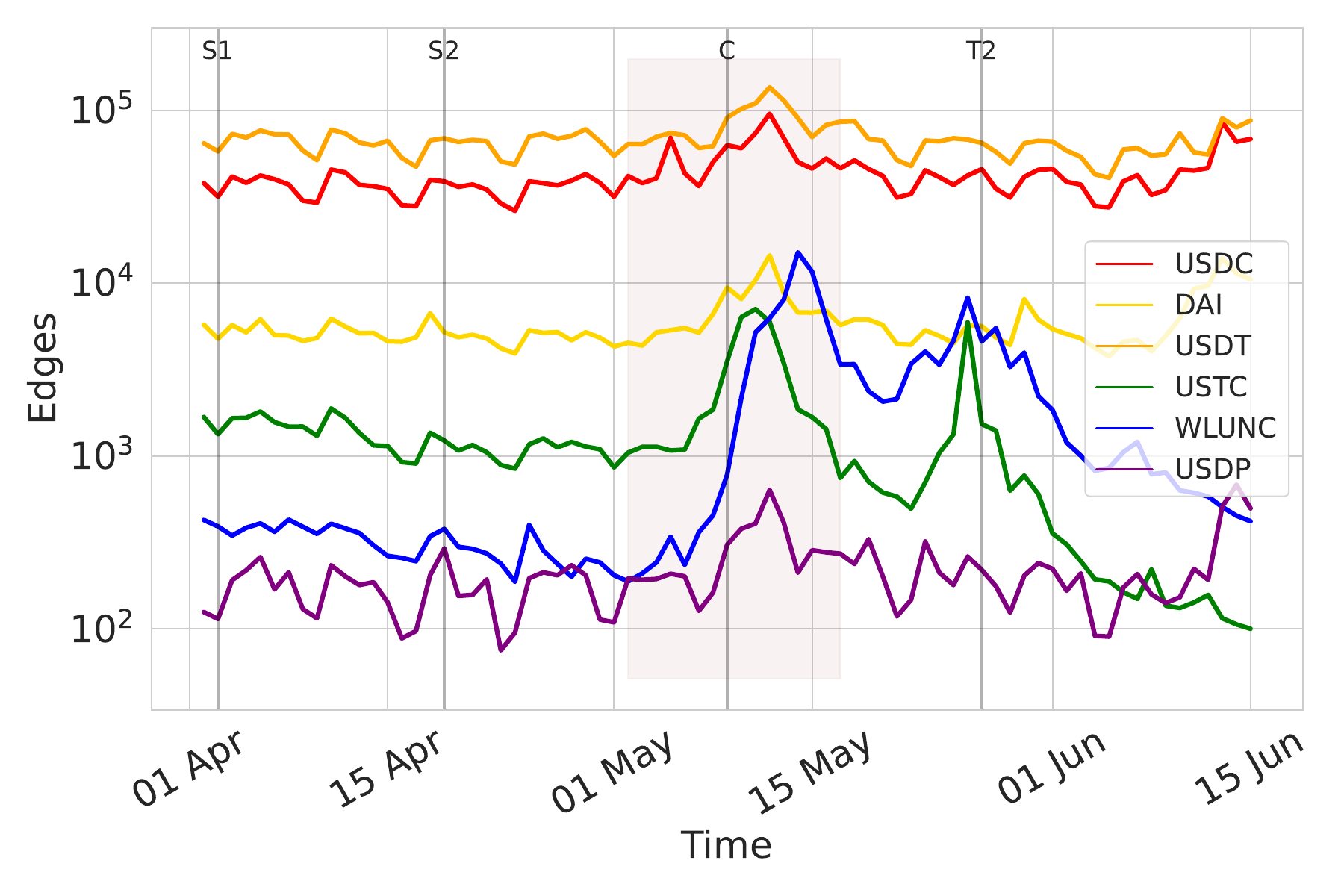}}
    \subcaption{Edges each day}
    \label{fig:Edges-1 day rolling}
    \end{minipage}
    \begin{minipage}[t]{0.48\textwidth}
    \centerline{\includegraphics[width=0.99\textwidth]{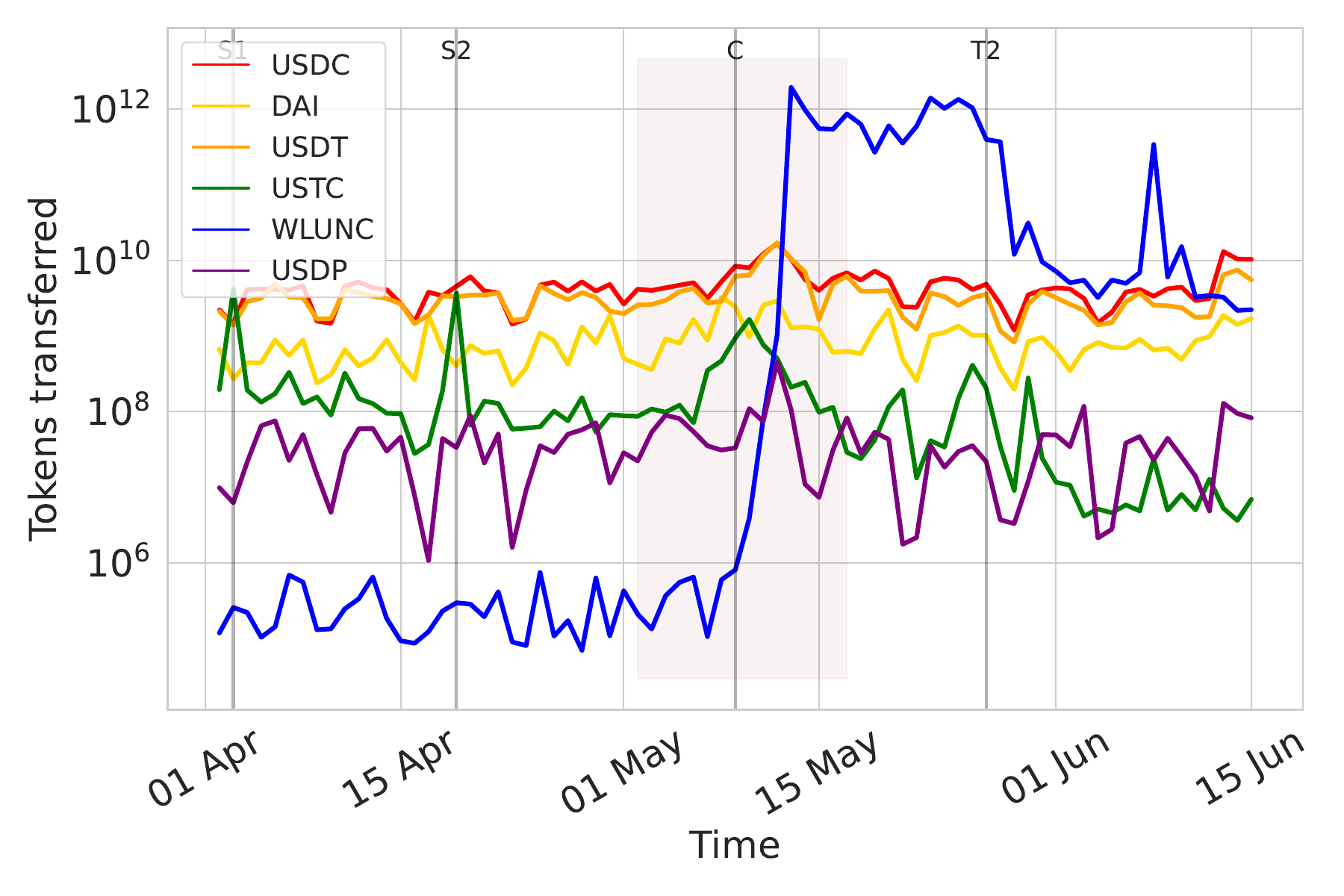}}
    \subcaption{Tokens transferred/traded each day}
    \label{fig:Selling-1 day rolling}
    \end{minipage}
    \begin{minipage}[t]{0.48\textwidth}
    \centerline{\includegraphics[width=0.99\textwidth]{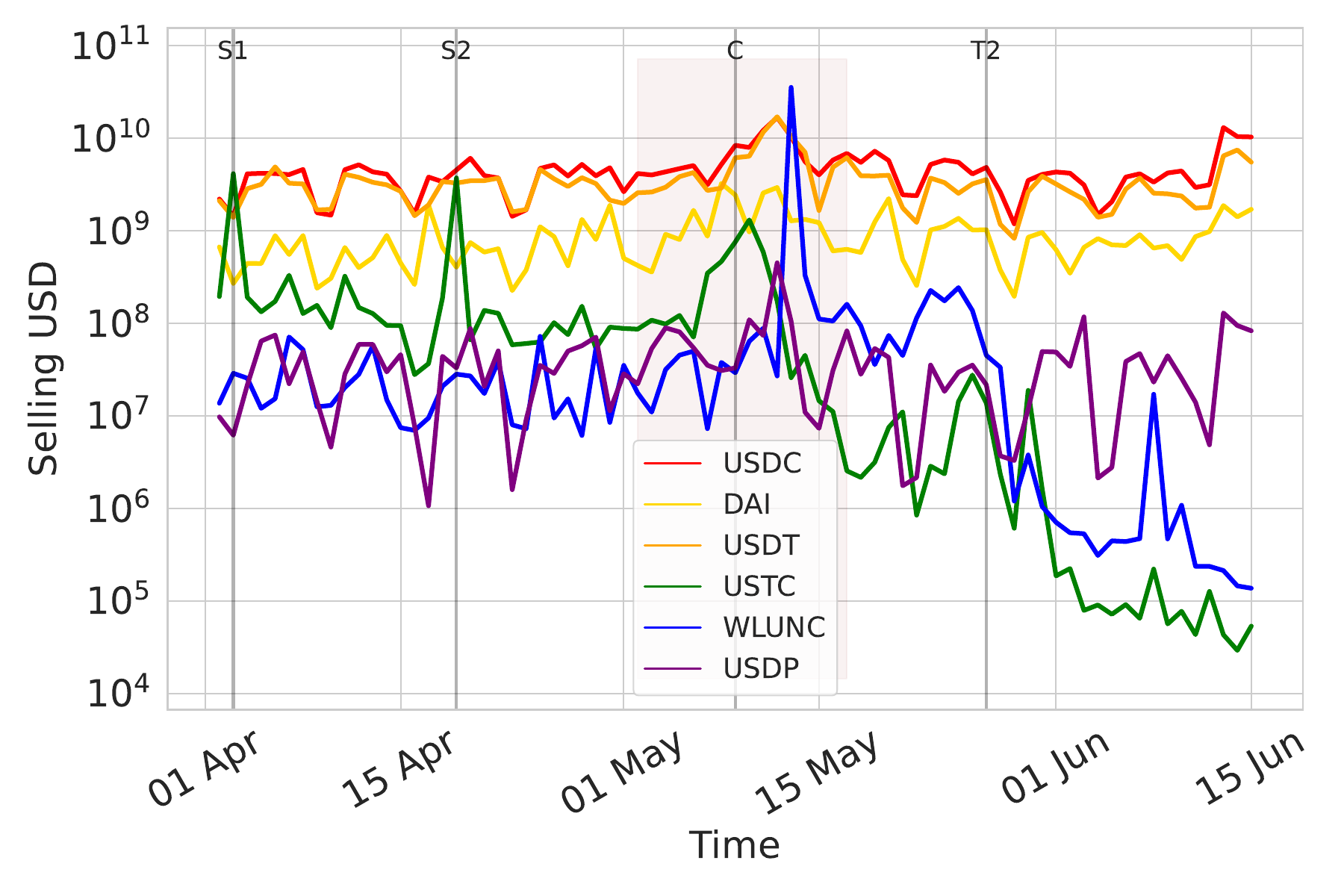}}
    \subcaption{Tokens transferred/traded (USD value) each day}
    \label{fig:Selling-USD-1 day rolling}
    \end{minipage}
\caption{Fluctuations in trading volume where each line represents one of the currencies under study. The number of transactions is the total transactions that day. The number of edges is the number of unique pairs of buyers transacting. The tokens transferred are the raw number of tokens and the value in USD uses the close price of that day's trading to convert this.}
\label{fig:str-1day-rolling}
\end{figure}

In \figurename\ref{fig:str-1day-rolling}, various aspects of the studied currencies are monitored. Firstly, we examine the daily number of transactions in \figurename~\ref{fig:Tnx-Count-1 day rolling}. The currencies exhibit different trading volumes, reflective of the size of their active user bases. When observing changes over time, we note the instability in values, with clear peaks visible for some currencies during the crash. Notably, currencies directly involved in the crash, WLUNC and USTC, demonstrate larger variations during this period. Post-crash, both currencies show a downward trend, although USTC displays a secondary peak in transaction volume. Similar observations are made when considering the number of edges in \figurename~\ref{fig:Edges-1 day rolling}. Pre-crash, we observe a lack of anomalous behaviour, followed by a surge during the crash period for WLUNC and USTC. Post-crash, WLUNC and USTC experienced a sudden increase in links to other users, interrupting a declining trend in transaction volume. Analyzing token movements in terms of total token transfers, presented in \figurename~\ref{fig:Selling-1 day rolling}, reveals noteworthy patterns. Before the crash, two anomalous peaks in USTC token transfers were observed on the 3rd April 2022 and the 19th April 2022 (labelled in the plot as S1 and S2, respectively). Other layers do not exhibit significant movements before the crash. The layers most influenced by the crash are primarily WLUNC and USTC, with WLUNC tokens experiencing a substantial increase in movements. Post-crash, the number of WLUNC tokens transferred remains high but eventually decreases. This decline can be attributed to the decreasing USD value of the tokens, as depicted in \figurename~\ref{fig:Selling-USD-1 day rolling}. Despite a consistent number of tokens transferred, their value diminishes over time. Other layers display relative stability, with a slight rise in USDC, USDT, and USDP token movements during the crash. In summary, suspicious anomalies can be observed before the crash, particularly in USTC token transfers.

\subsection{What are the effects of the attack on the system beyond the most directly affected currencies?}
\begin{figure}[ht!] 
    \begin{minipage}[t]{0.48\textwidth}
    \centerline{\includegraphics[width=0.9\textwidth]{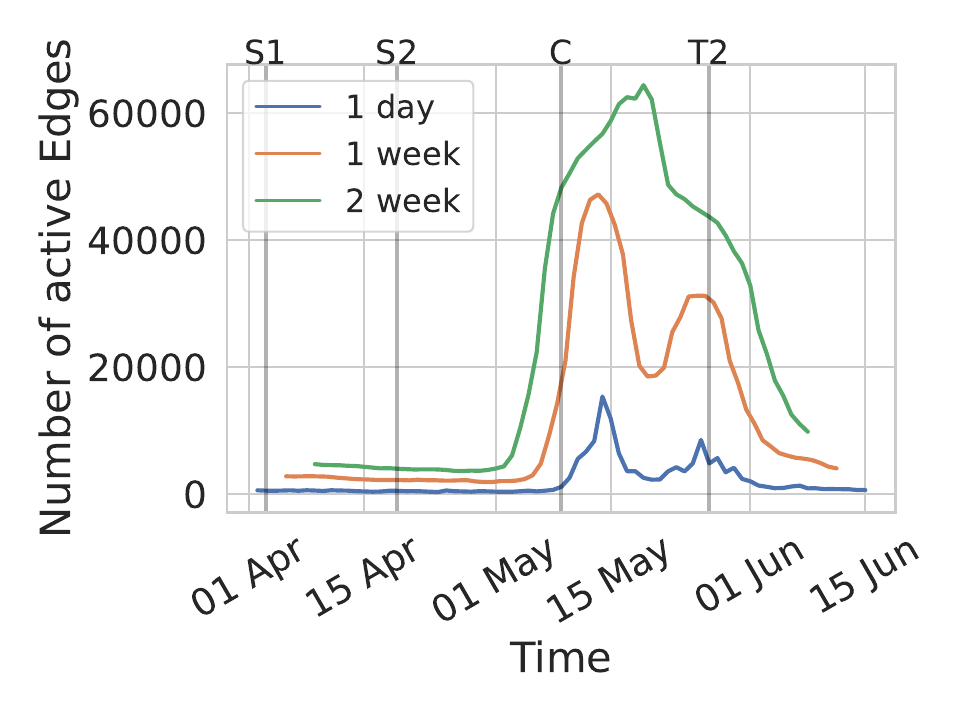}}
    \subcaption{New Edges in WLUNC layer}
    \label{fig:LUNC-new-Edges}
    \end{minipage}
    \begin{minipage}[t]{0.48\textwidth}
    \centerline{\includegraphics[width=0.9\textwidth]{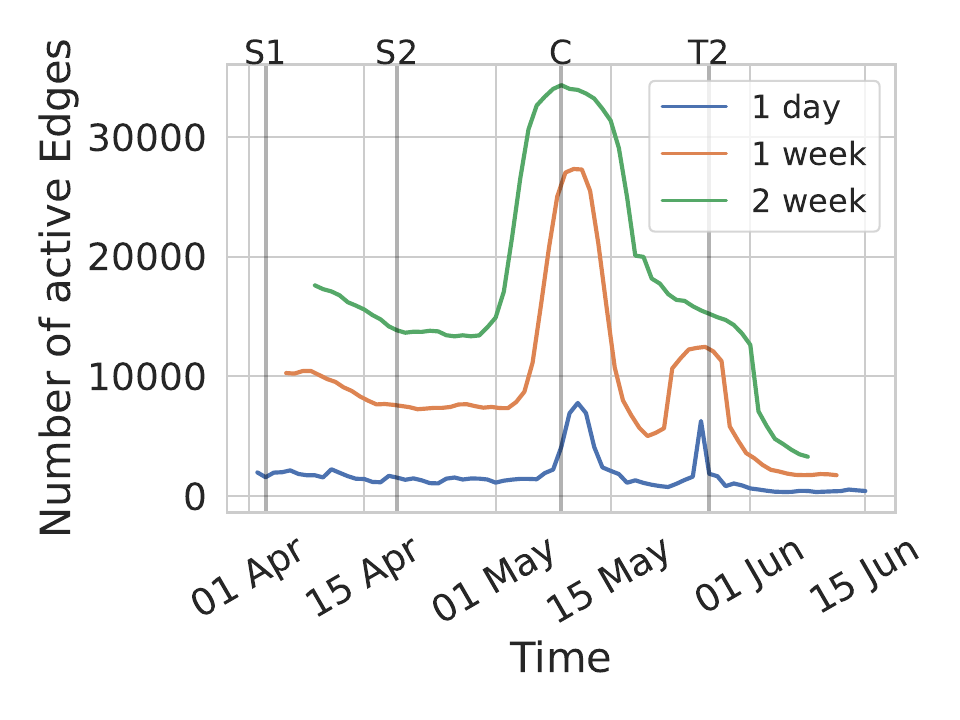}}
    \subcaption{New Edges in USTC layer}
    \label{fig:USTC-new-Edges}
    \end{minipage}
    \begin{minipage}[t]{0.48\textwidth}
    \centerline{\includegraphics[width=0.9\textwidth]{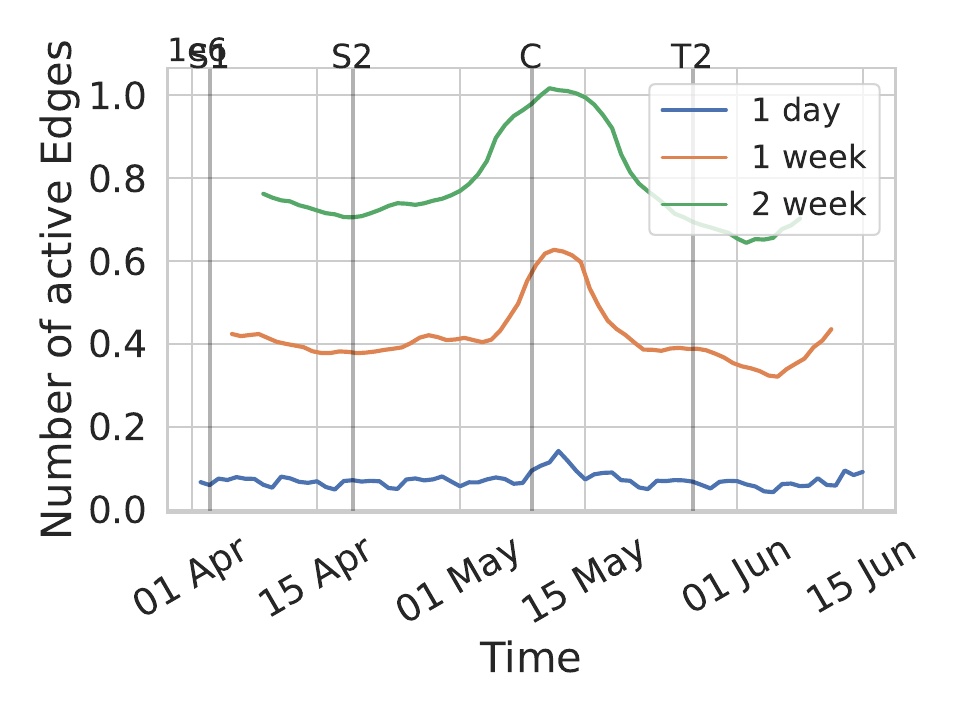}}
    \subcaption{New Edges in USDT layer}
    \label{fig:USDT-new-Edges}
    \end{minipage}
\caption{Monitoring the new edges between unique user pairs across different window sizes. Here {\it new} means that this is the first window in which they are present. On the X-axis: Time. On the Y-axis: number of edges according to the selected time window.}
\label{fig:new-edges}
\end{figure}

Our objective is to examine the driving forces behind the system, discerning whether it is propelled by returning users who consistently contribute or by new users who have not been encountered previously. To achieve this, we adopt a methodology inspired by prior research ~\cite{arnold2021moving}, wherein distinctions are made between new users or edges within a specified window. These entities encompass users and pairwise interactions occurring within the window and have not been active at any time prior to the selected window. The adjustment of the window size $\tau$ enables varied definitions of an active user. We monitor, for each layer, the number of new interactions according to daily, weekly, and bi-weekly windows. An edge is considered new if it is the first time window it occurs in. 

We show the main currencies WLUNC and USTC, in Figures ~\ref{fig:LUNC-new-Edges} and ~\ref{fig:USTC-new-Edges} respectively. In both currencies, an intriguing pattern emerges. There are two peaks in the number of new edges, with the first peak coinciding with the crash period. This is likely a result of currency holders responding by engaging in trades with new users (disposing of the coin) while new users enter the system, likely driven by speculative motives. The second notable surge in USTC and WLUNC plots aligns with the launch of Terra 2.0 discussed in Section \ref{sec:background}. A plausible explanation for this second surge stems from the criteria for distributing new LUNA tokens on Terra 2.0 \cite{news_proposal_1623_terra_2022,medium_mc_terra_2022}. The proposal specifies eligibility criteria for users to receive tokens, based on selected snapshots pre-attack and post-attack. The Governance chose 7th May 2022 at 11 pm for the pre-crash snapshot and 27th May 2022 at 3.59 AM for the post-crash snapshot. Notably, this choice favours the users who were invested in Terra or Luna before the crash. With the voting concluding on 25th May 2022, making the plan official, and the target launch date on 27th May 2022, users who sold during the crash but intended to join Terra 2.0 had an incentive to purchase WLUNC tokens before the 27th May 2022 deadline. This potentially explains the spikes in USTC and WLUNC on 26th May 2022.

By contrast, other currencies exhibit different behaviour. Their plots resemble that of USDT in \figurename~\ref{fig:USDT-new-Edges}, with most currencies showing only one peak immediately following the crash. While the crash induces a behaviour change, the launch of Terra 2.0 does not significantly impact the other layers.

\subsection{Can the after-effects of the attack be seen on graph structure as a whole?}

\begin{figure}[ht!] 
    \begin{minipage}[t]{0.48\textwidth}
    \centerline{\includegraphics[width=0.7\textwidth]{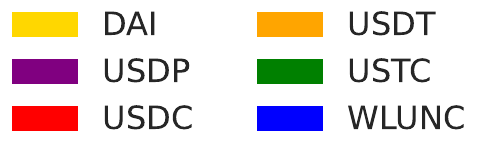}}
    \subcaption{Considered currencies (legend)}
    \label{fig:legend-currencies}
    \end{minipage}
    \begin{minipage}[t]{0.48\textwidth}
    \centerline{\includegraphics[width=0.9\textwidth]{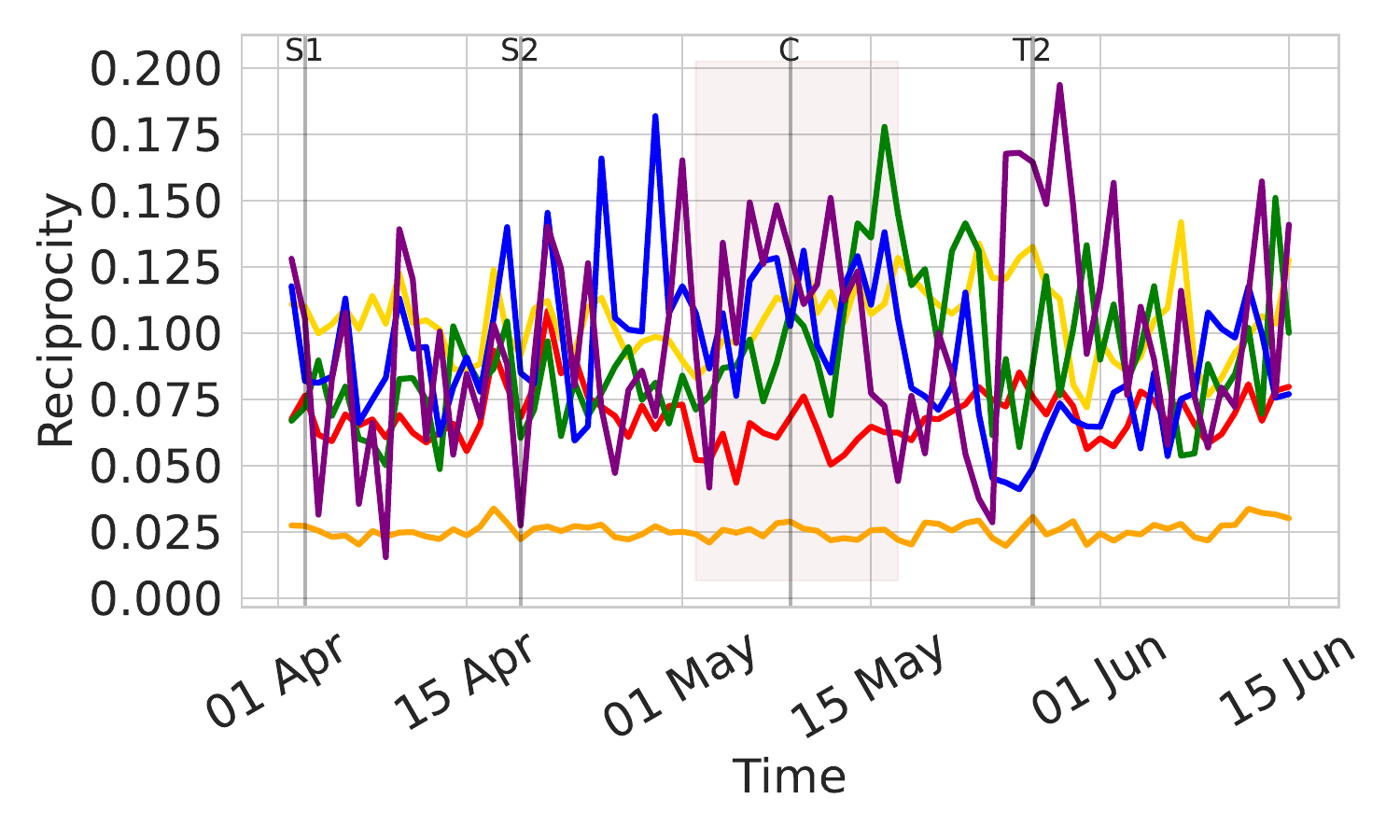}}
    \subcaption{Reciprocity values}
    \label{fig:Sub-active-reciprocity-raph-window-1 day-step-1 day}
    \end{minipage}
    \begin{minipage}[t]{0.48\textwidth}
    \centerline{\includegraphics[width=0.9\textwidth]{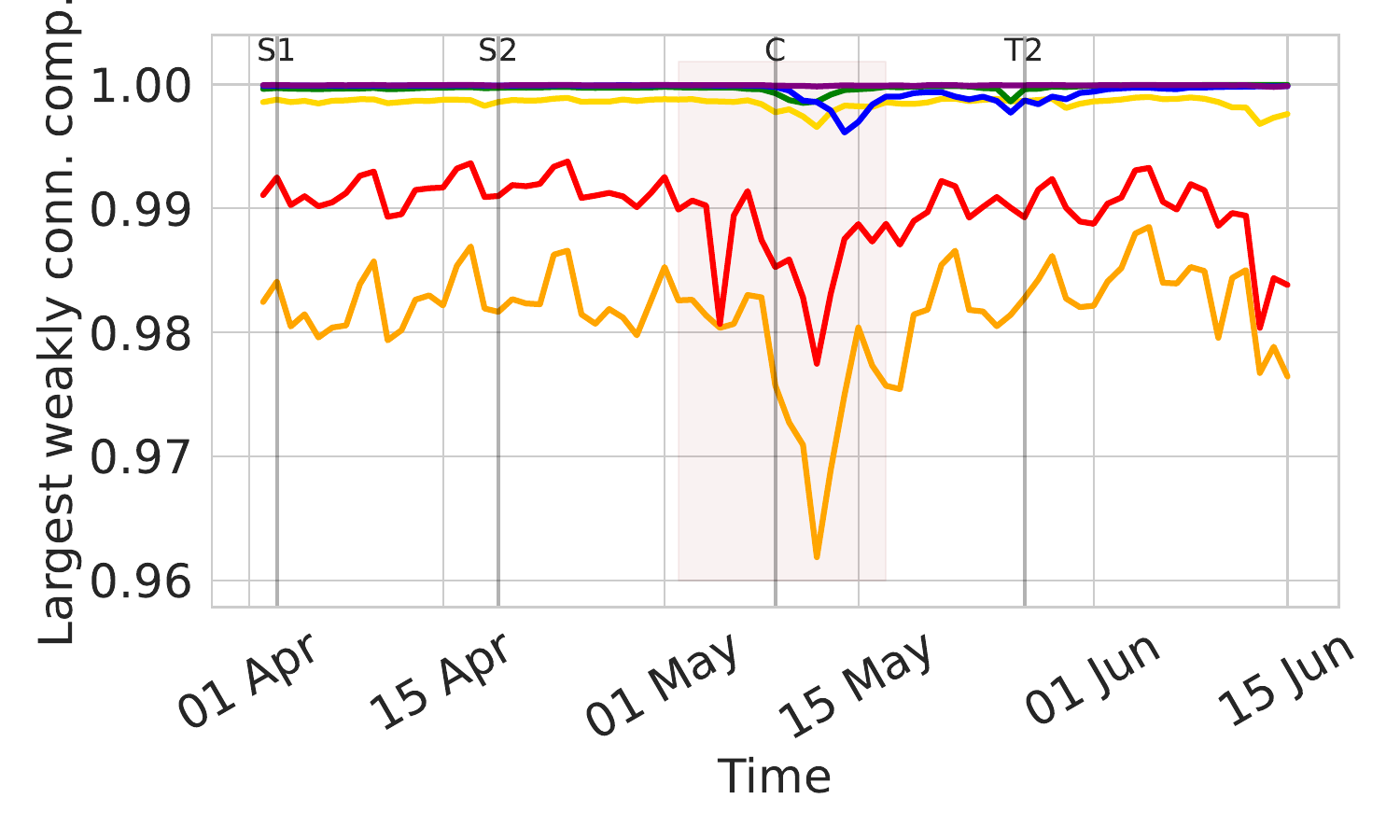}}
    \subcaption{Fraction of users in the largest weakly connected component}
    \label{fig:Sub-active-largest-wcc-raph-window-1 day-step-1 day}
    \end{minipage}
    \begin{minipage}[t]{0.48\textwidth}
    \centerline{\includegraphics[width=0.9\textwidth]{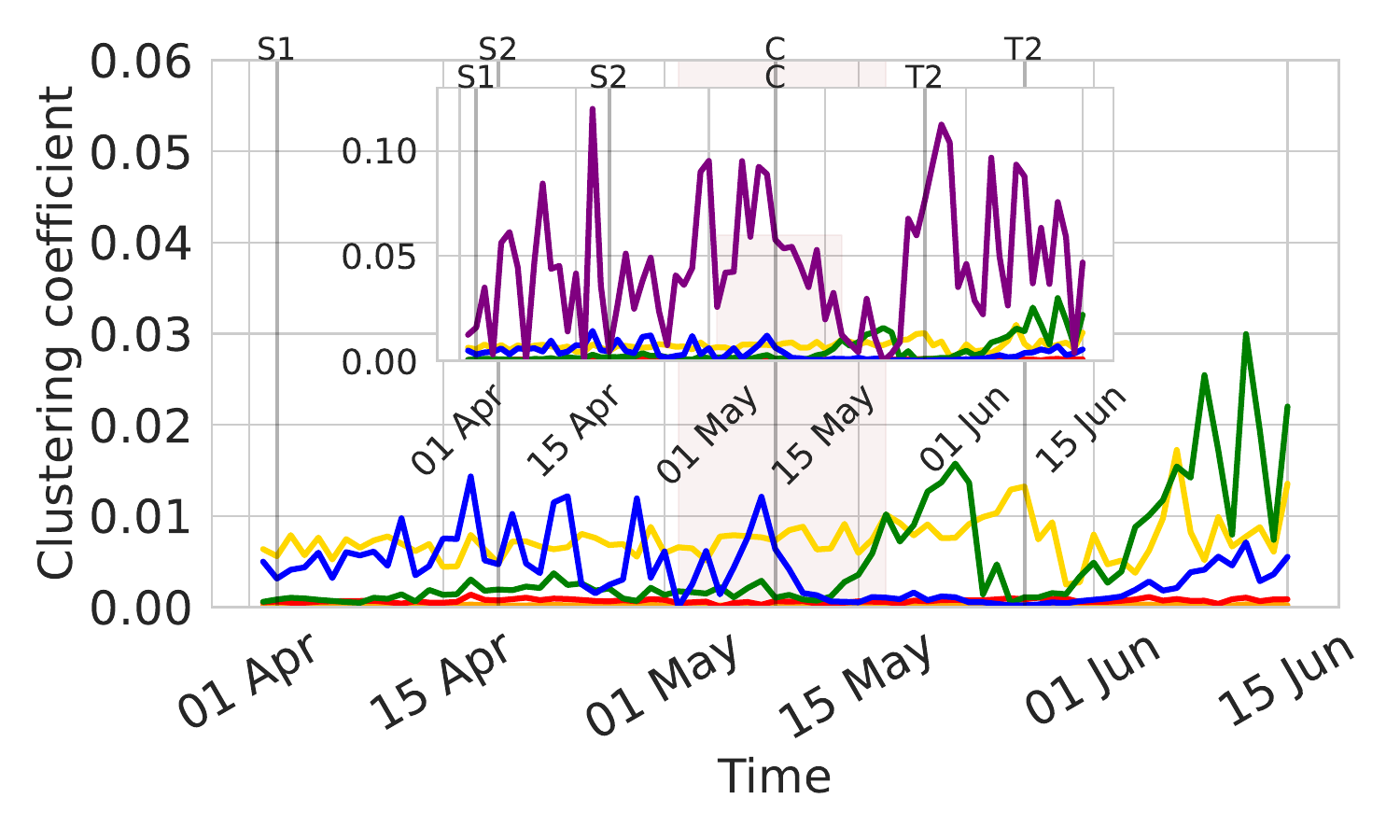}}
    \subcaption{Clustering coefficient}
    \label{fig:Sub-active-clustering-coeff-raph-window-1 day-step-1 day}
    \end{minipage}
    \begin{minipage}[t]{0.48\textwidth}
    \centerline{\includegraphics[width=0.9\textwidth]{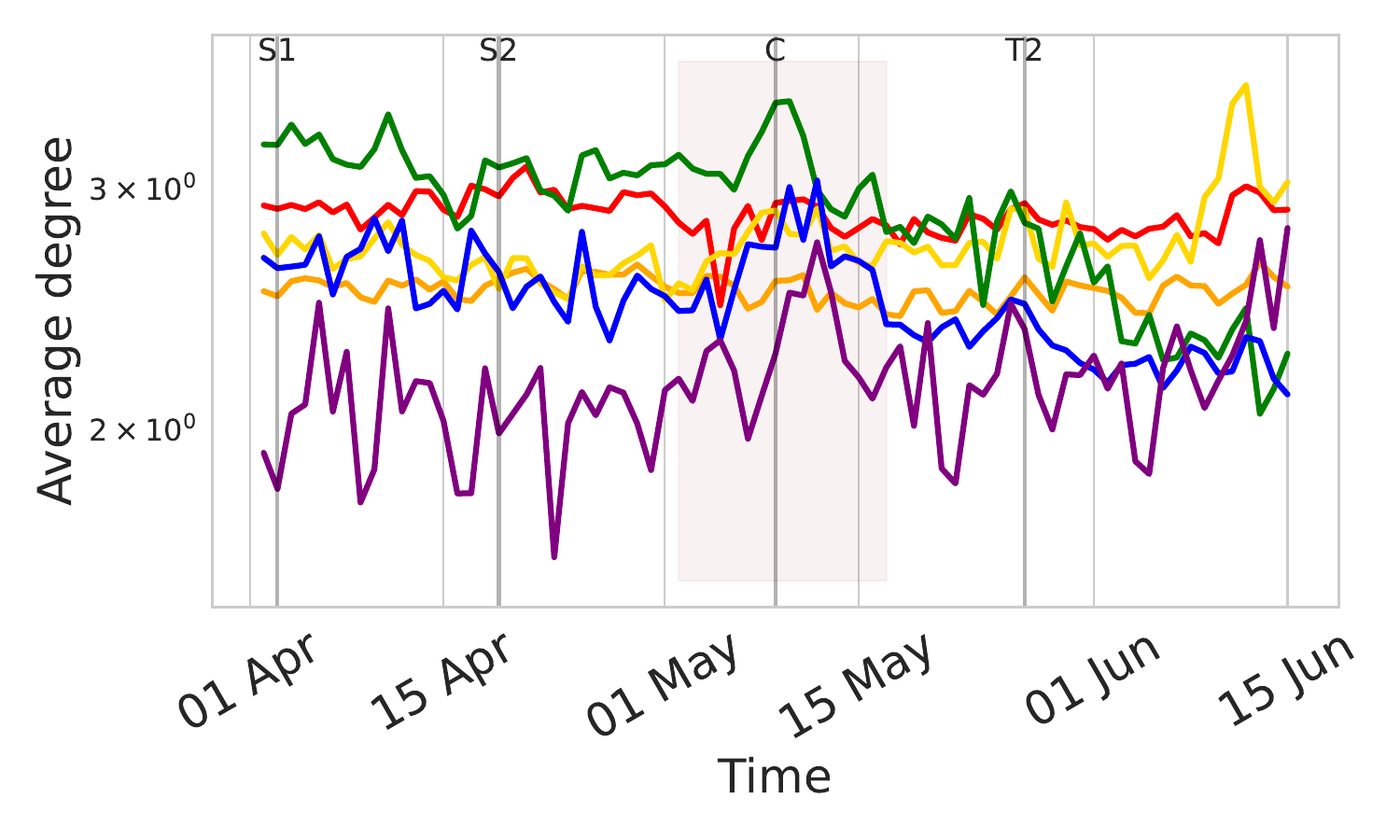}}
    \subcaption{Average Degree}
    \label{fig:Sub-active-average-degree-raph-window-1 day-step-1 day}
    \end{minipage}
    \begin{minipage}[t]{0.48\textwidth}
    \centerline{\includegraphics[width=0.9\textwidth]{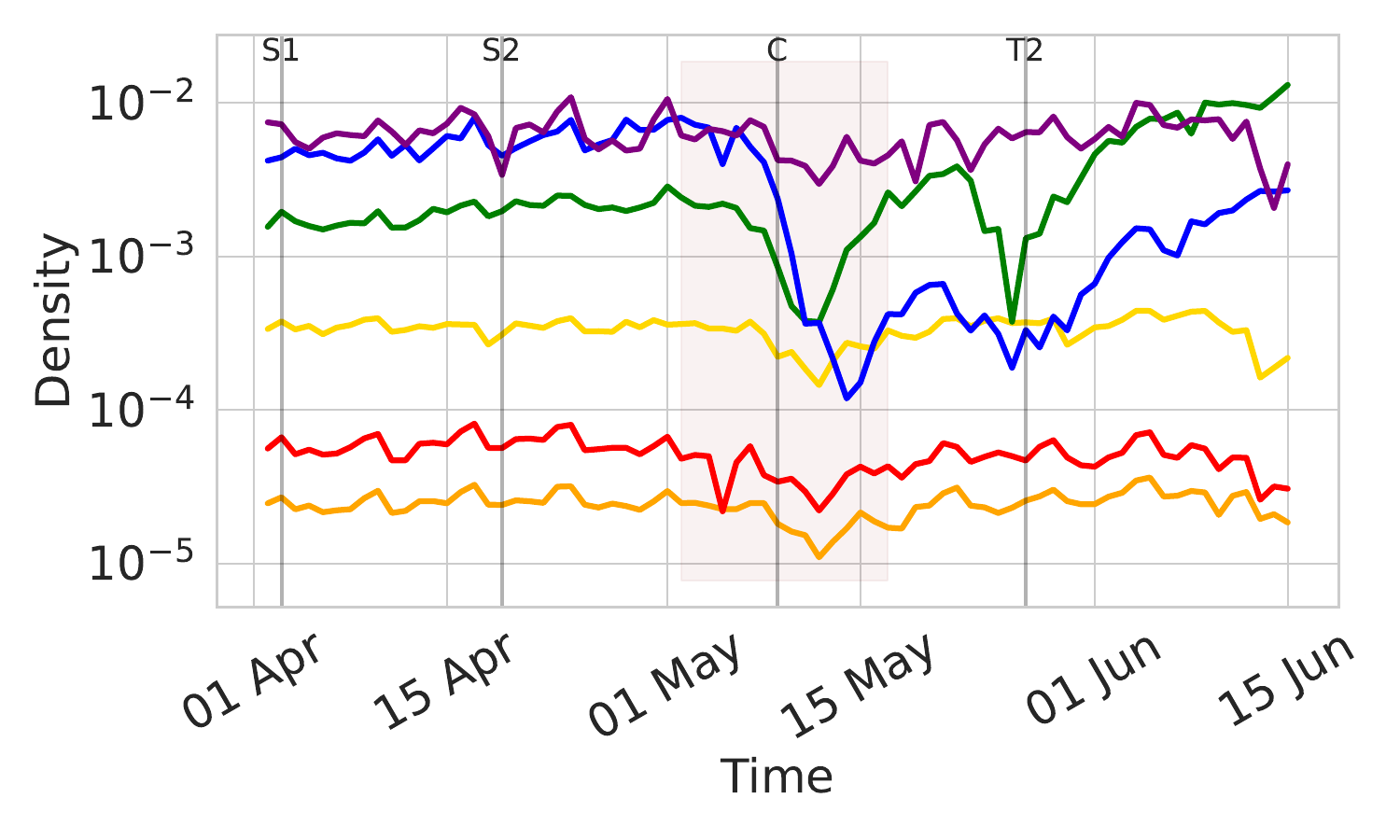}}
    \subcaption{Density}
    \label{fig:Sub-active-density-raph-window-1 day-step-1 day}
    \end{minipage}
\caption{Graph measures over time. We monitor different graph measures over time, For each day, the interactions on each currency are treated as an individual graph. On the X-axis: Time. On the Y-axis: the selected graph measures values over time. For clustering coefficient (c), one of the layers exhibits outlier values, so we focus on the main currencies in the main plot, while we provide the complete Figure as an inset.}
\label{fig:metrics-rolling}
\end{figure}

We hypothesize that a crash could lead to changes to the graph structure. Therefore, we delved into the presence of short and long-term trends using traditional graph measures, including reciprocity, clustering coefficients, weakly connected components, mean degree, and graph density. The analysis was conducted layer by layer, treating each currency as an individual graph, and the results are presented in \figurename~\ref{fig:metrics-rolling}.

The reciprocity measure (see \figurename~\ref{fig:Sub-active-reciprocity-raph-window-1 day-step-1 day}) remains relatively stable for all currencies. Notable spikes are observed on the WLUNC Layer before the crash, and on the USTC layer, a rising trend is evident during the crash. However, post-crash values return to a range similar to pre-crash levels. The other currencies do not exhibit significant changes in their reciprocity levels. Examining the size of the largest weakly connected component (see \figurename~\ref{fig:Sub-active-largest-wcc-raph-window-1 day-step-1 day}), we observe a drop in the days right after the crash (labelled C), impacting all layers, with notable shifts in the USDC and USDT layers. However, post-crash values revert to the pre-crash range. A small change is noted preceding the launch of Terra 2.0 (labelled T2). The density measure (see \figurename~\ref{fig:Sub-active-density-raph-window-1 day-step-1 day}) shows a significant drop right after the crash (labelled C) for Luna and USTC layers, likely influenced by an increasing number of active users during that time. Density values recover after the crash. Additional changes in density values align with the launch of Terra 2.0, particularly the rise on the USTC layer. The other currencies exhibit similar behaviour, with a drop in density values during the crash; but are characterised by relatively stable density values throughout the observation period. The clustering coefficient measure (see \figurename~\ref{fig:Sub-active-clustering-coeff-raph-window-1 day-step-1 day}) showcases fewer changes. On the WLUNC layer, there's a drop after the crash, while the USTC layer sees a rise in the last days of the crash, with values remaining high post-crash. The other layers do not experience significant changes. Similarly, the mean degree (Figure \ref{fig:Sub-active-average-degree-raph-window-1 day-step-1 day}) does not highlight substantial changes, except for temporary rises during the crash, aligning with the overall increase in activity.

In summary, certain graph measures confirm that the graph structure undergoes changes during the observation period, especially during the crash. However, users who remain active tend to re-establish a similar graph structure after the disruptive events pass.

\subsection{Where did user interest move post crash?}

\begin{figure}[ht!] 
    \begin{minipage}[t]{0.48\textwidth}
    \centerline{\includegraphics[width=0.9\textwidth]{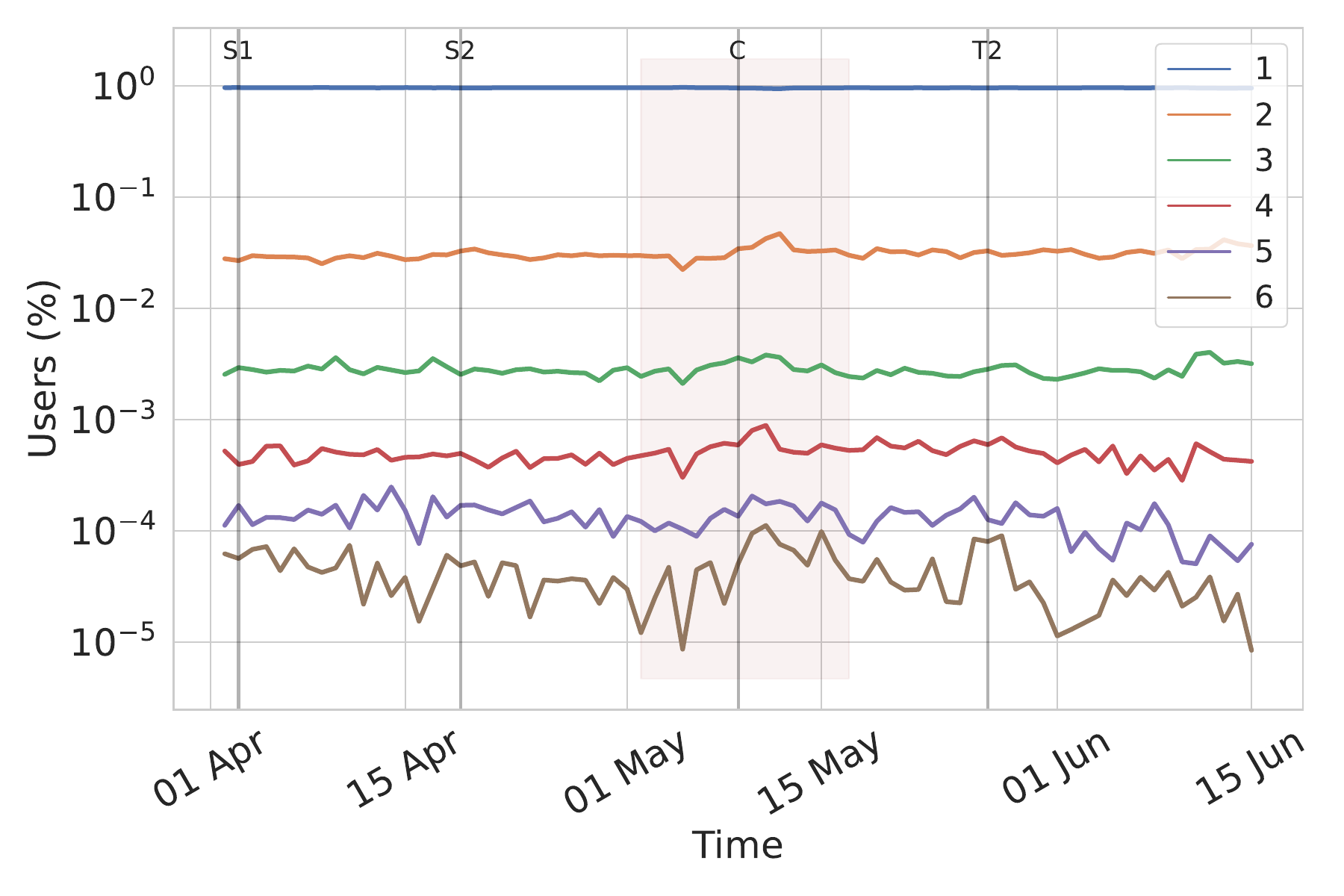}}
    \subcaption{Number of layers for all users}
    \label{fig:Multilayer-act-1}
    \end{minipage}
    
    \begin{minipage}[t]{0.48\textwidth}
    \centerline{\includegraphics[width=0.9\textwidth]{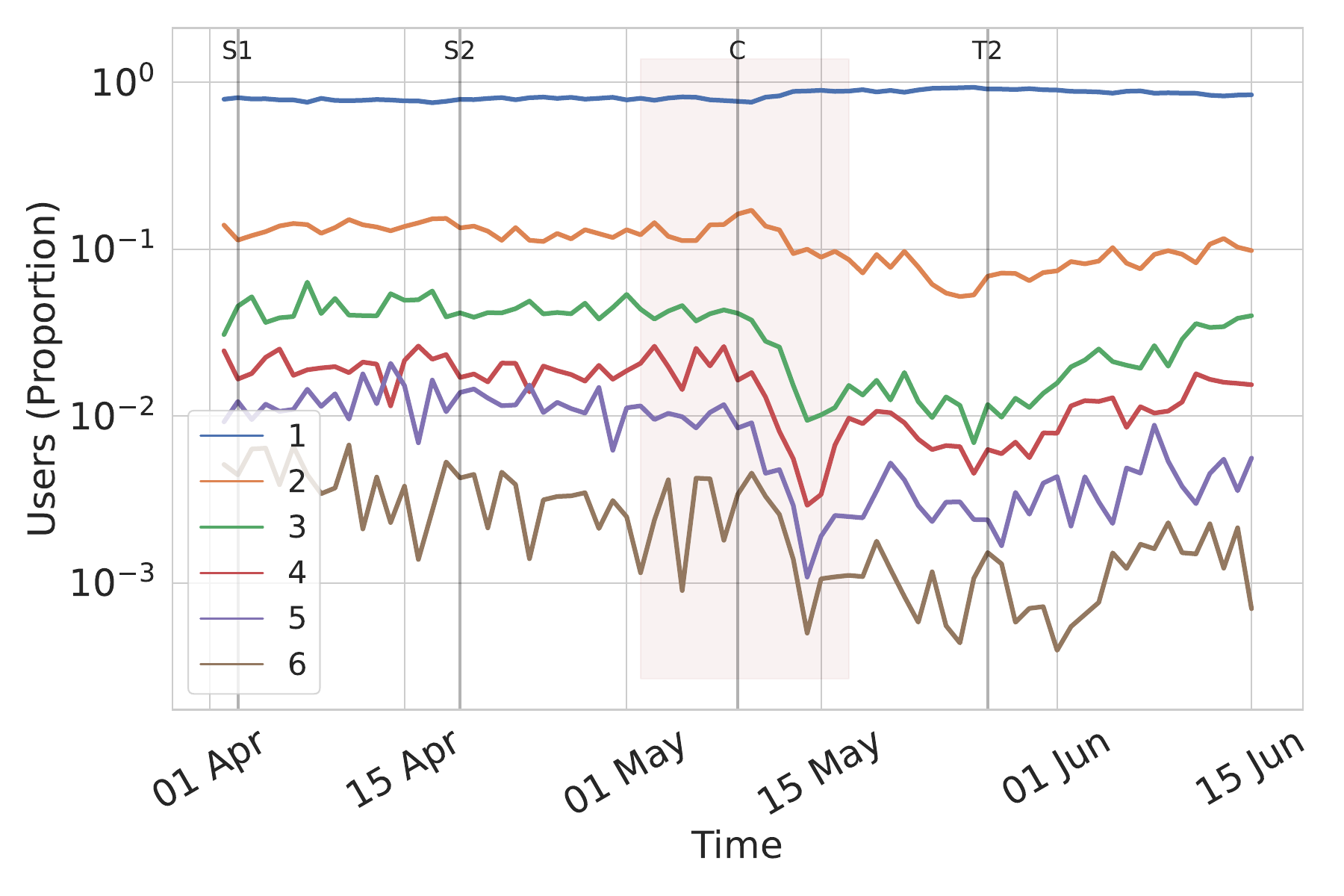}}
    \subcaption{Number of layers (users that use WLUNC)}
    \label{fig:Multilayer-act-focus-LUNC-1}
    \end{minipage}
    \begin{minipage}[t]{0.48\textwidth}
    \centerline{\includegraphics[width=0.9\textwidth]{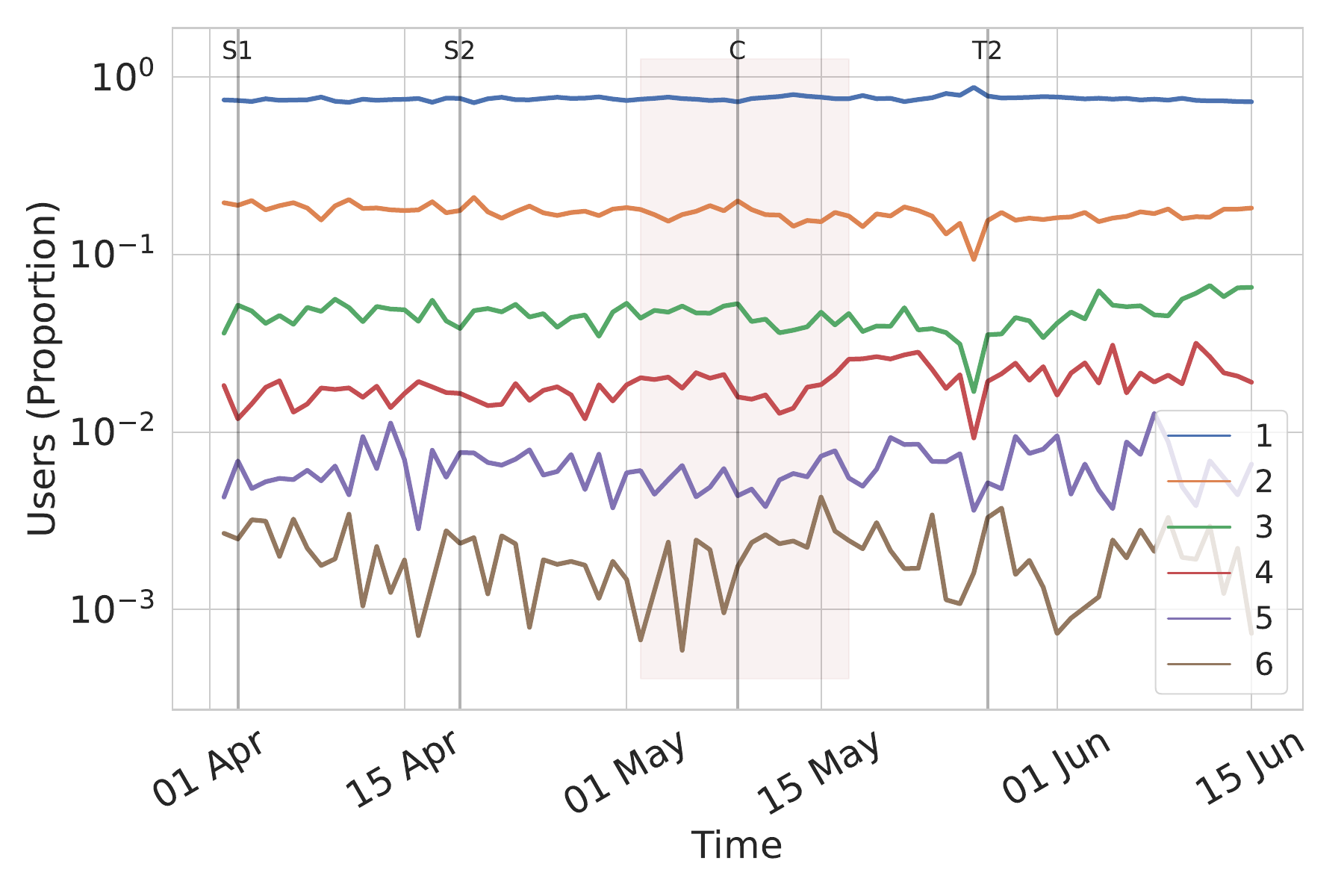}}
    \subcaption{Number of layers (users that use USTC) }
    \label{fig:Multilayer-act-focus-USTC-1}
    \end{minipage}
\caption{Number of layers in which users are active. At each timestep, we check the number of layers in which the users are active, giving them a label (1...6). Each line represents the changes in the possible cases (1...6). On the X-axis: Time. On the Y-axis: fraction of each possible value.}
\label{fig:multilayer-act}
\end{figure}

We might expect users to react to the crash in WLUNC and USTC by diversifying into other currencies. \figurename~\ref{fig:multilayer-act} shows the number of layers that users are present in (equivalently to the number of currencies they trade that day). For all users (see \figurename~\ref{fig:Multilayer-act-1}) we can see very little difference but perhaps a small increase immediately after the centre of the crash (label C). If we look at only those users who trade WLUNC (see \figurename~\ref{fig:Multilayer-act-focus-LUNC-1}) we can see the opposite behaviour a decrease in the number trading multiple currencies (the apparently smaller increase in the number of trades of a single currency is caused by the log scale). The likely cause is the large increase in users trading WLUNC on those days who are simply trying to sell their WLUNC and are not interested in other trades. A similar but less pronounced trend is seen in the graph of users who trade USTC (see \figurename~\ref{fig:Multilayer-act-focus-USTC-1}) and this figure also shows a similar trend around event $T2$.

\begin{figure}[ht!] 
    \begin{minipage}[t]{0.48\textwidth}
    \centerline{\includegraphics[width=0.9\textwidth]{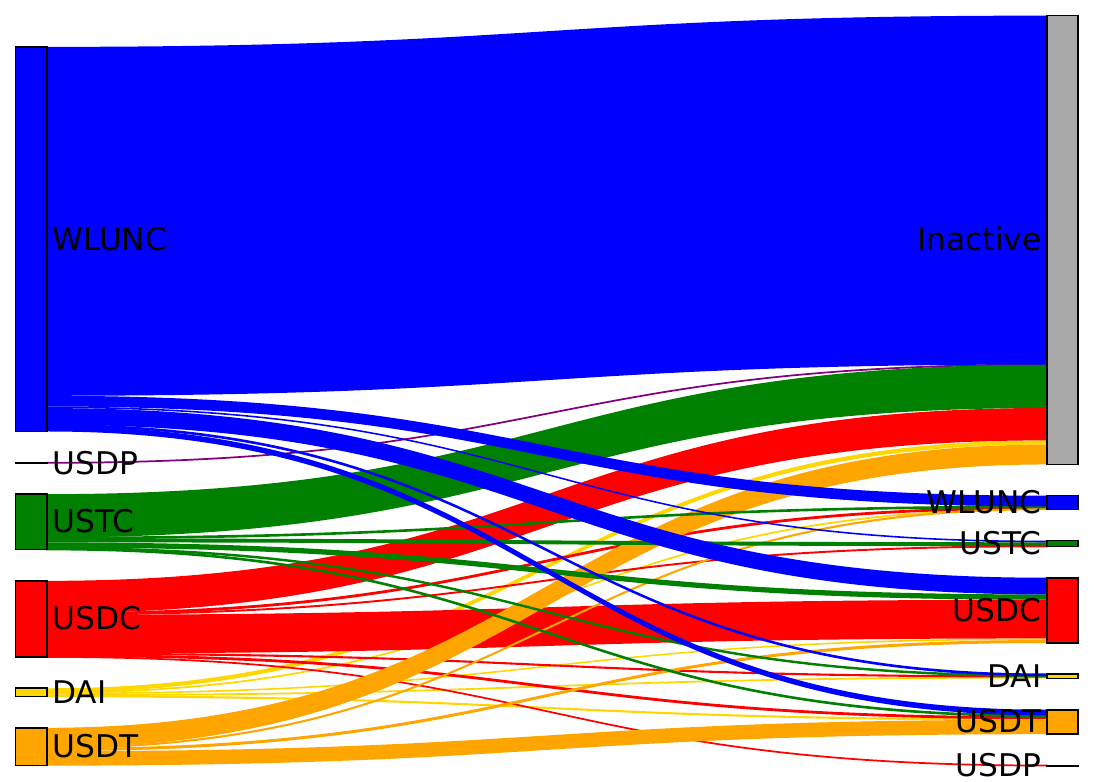}}
    \subcaption{Favourite layer shift for users active on WLUNC}
    \label{fig:sankey-table-fav-prepost-Edges-LUNC}
    \end{minipage}
    \begin{minipage}[t]{0.48\textwidth}
    \centerline{\includegraphics[width=0.9\textwidth]{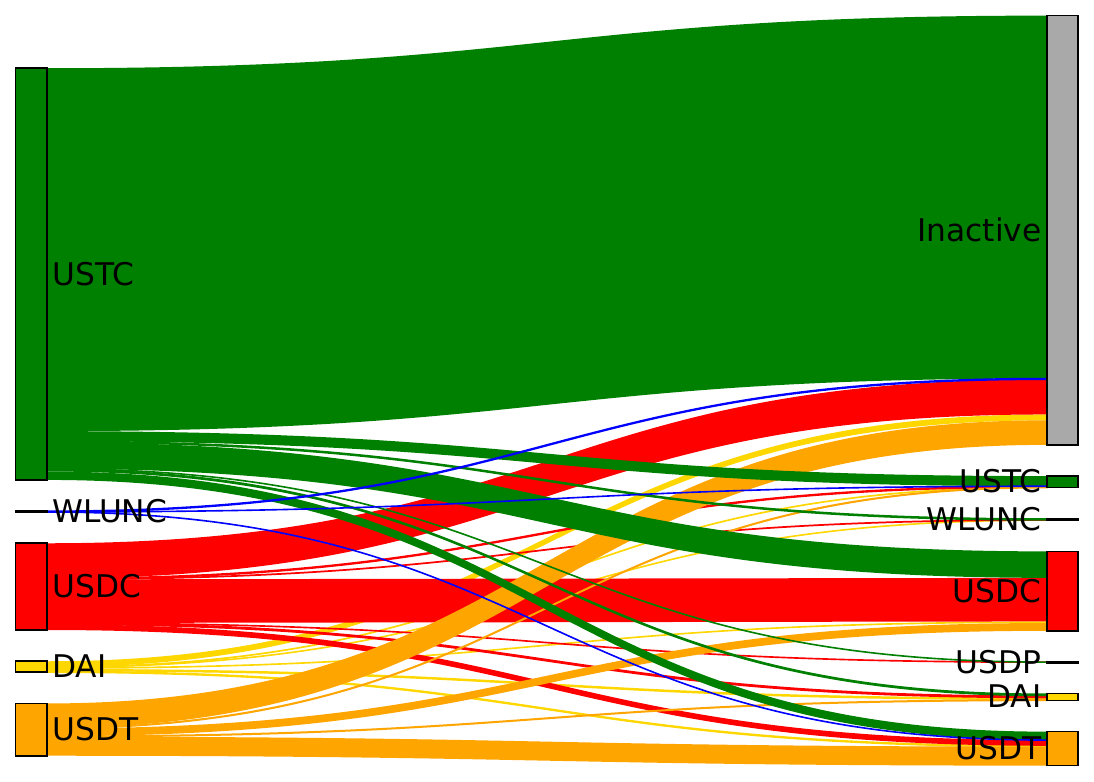}}
    \subcaption{Favourite layer shift for users active on USTC}
    \label{fig:sankey-table-fav-prepost-Edges-USTC}
    \end{minipage}
    \begin{minipage}[t]{0.48\textwidth}
    \centerline{\includegraphics[width=0.9\textwidth]{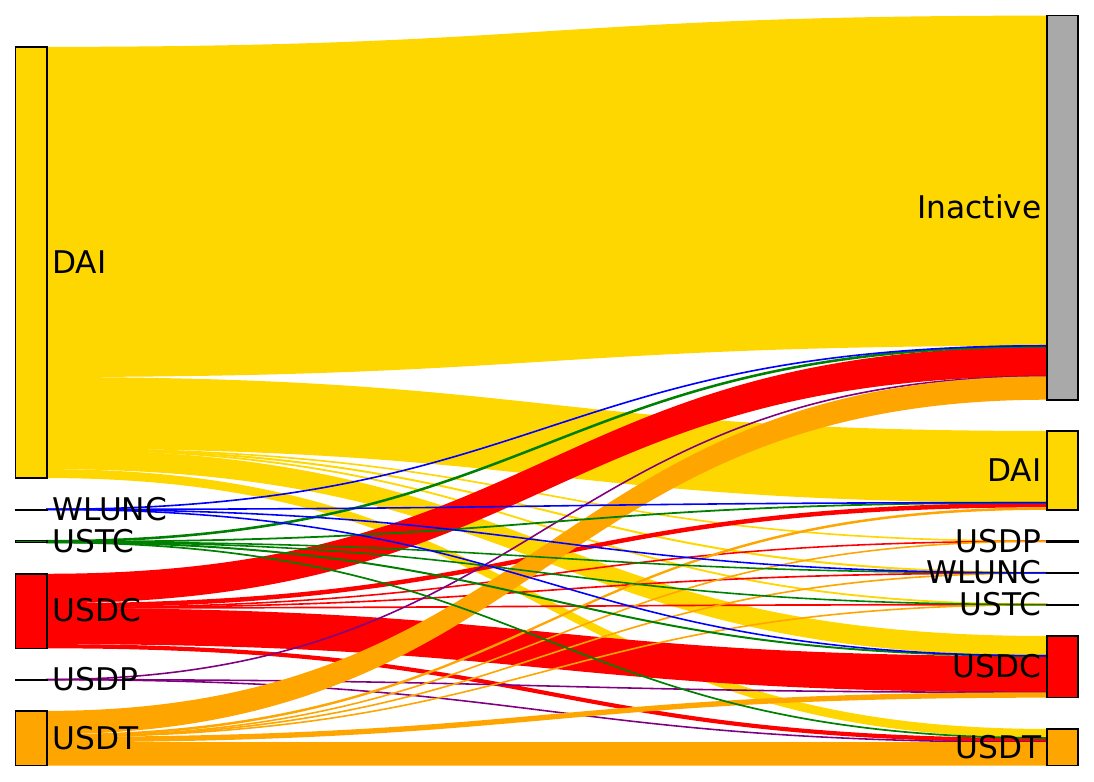}}
    \subcaption{Favourite layer shift for users active on DAI}
    \label{fig:sankey-table-fav-prepost-Edges-Dai}
    \end{minipage}
    \begin{minipage}[t]{0.48\textwidth}
    \centerline{\includegraphics[width=0.9\textwidth]{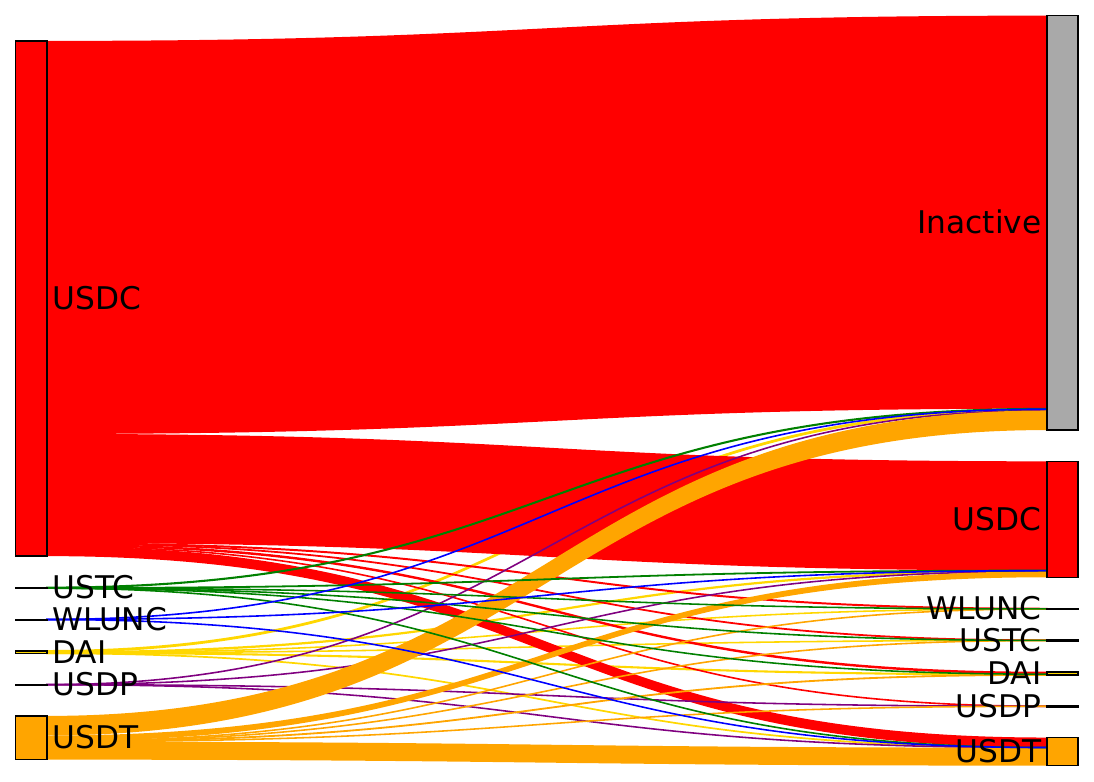}}
    \subcaption{Favourite layer shift for users active on USDC}
    \label{fig:sankey-table-fav-prepost-Edges-USD}
    \end{minipage}
    \begin{minipage}[t]{0.48\textwidth}
    \centerline{\includegraphics[width=0.9\textwidth]{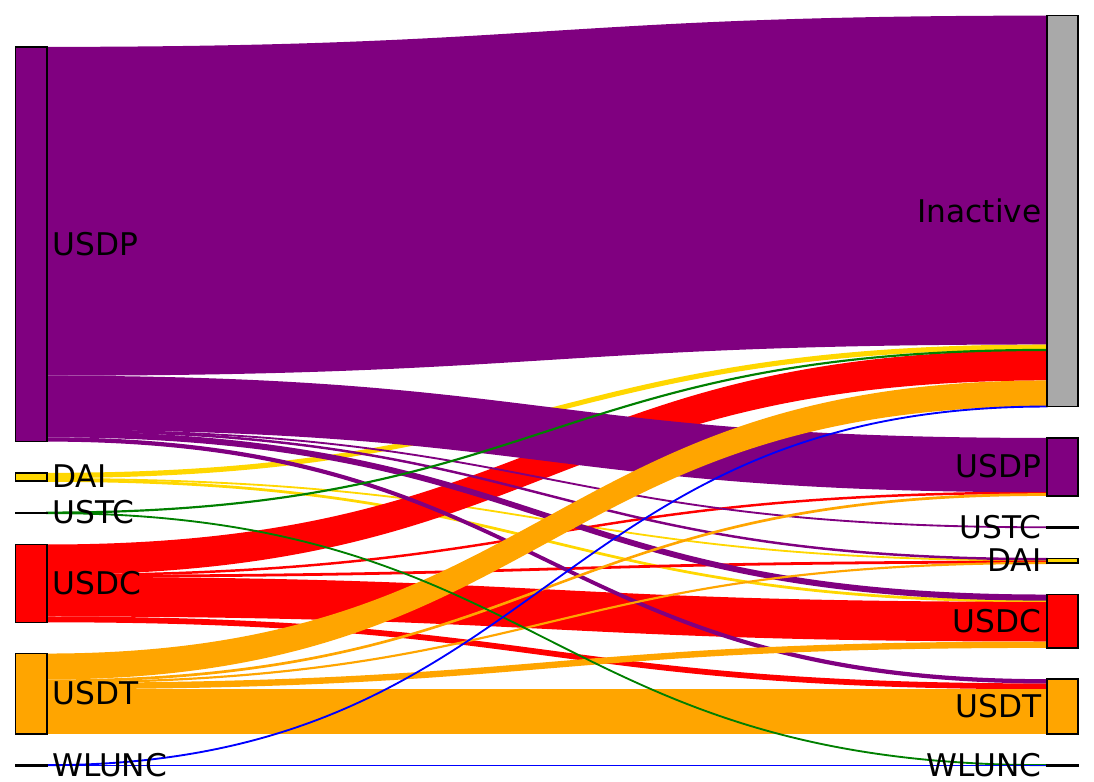}}
    \subcaption{Favourite layer shift for users active on USDP}
    \label{fig:sankey-table-fav-prepost-Edges-USDP}
    \end{minipage}
    \begin{minipage}[t]{0.48\textwidth}
    \centerline{\includegraphics[width=0.9\textwidth]{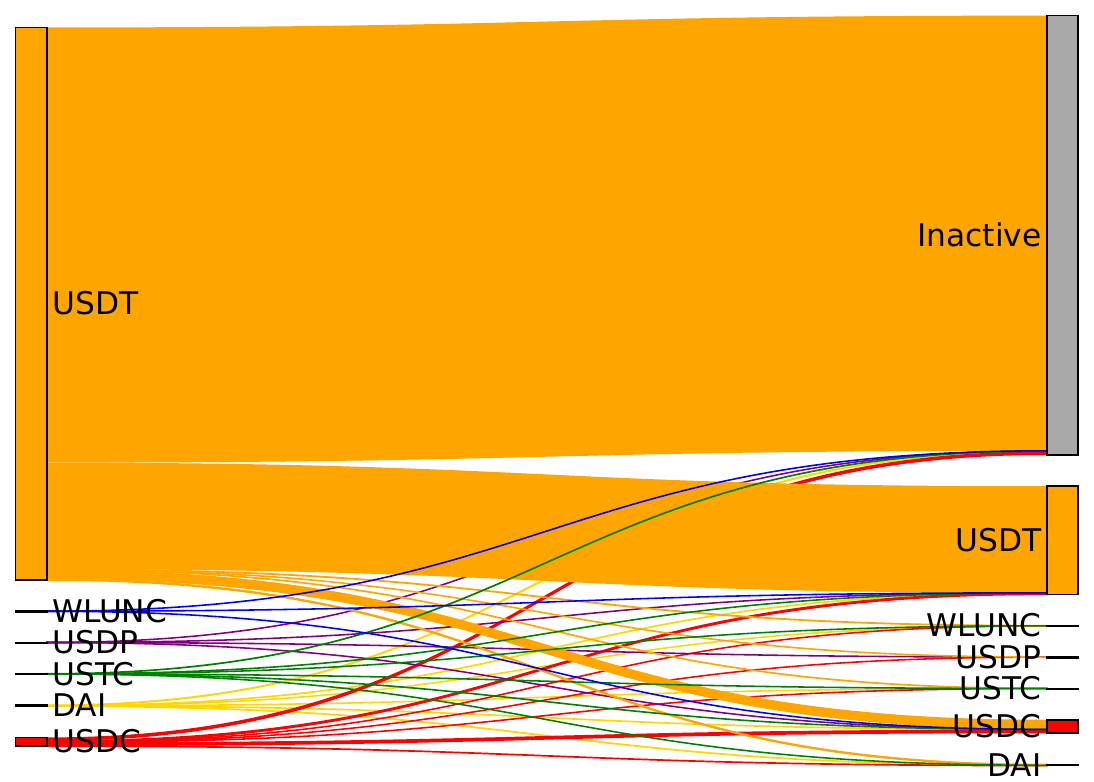}}
    \subcaption{Favourite layer shift for users active on USDT}
    \label{fig:sankey-table-fav-prepost-Edges-USDT}
    \end{minipage}
\caption{Changes of favourite layer for users selected pre-crash and observed post-crash. Each graph is focused on a layer: given a layer, we select users that had at least 1 transaction on that layer. A node is labelled with its favourite layer i.e. the layer in which they have the most Edges. }
\label{fig:flows-fav-layer}
\end{figure}

To gain a deeper understanding of how user behaviour evolved, we conducted an additional analysis focusing on users' favourite layers—indicating the layer where they are most active—before and after the crash. We presume that users may shift their attention to different currencies post-crash, especially away from the currencies most involved in the crash. The shifts are visualised using Sankey diagrams in \figurename~\ref{fig:flows-fav-layer}. The Sankey diagrams reveal that users active before the crash tend to be inactive after the crash. Across all currencies, among those who remain active, there is a notable trend of shifting focus to different currencies; however, there are differences among currencies. Specifically, in the currencies directly involved in the crash, Terra and Luna (see Figures \ref{fig:sankey-table-fav-prepost-Edges-USTC} and \ref{fig:sankey-table-fav-prepost-Edges-LUNC}) we can observe how among those who remain active, there is a notable trend of shifting focus to different currencies. For example, in the case of WLUNC, users predominantly migrate to USDC and USDT. Instead, on the other layers (DAI, USDC, USDT, USDP), a higher proportion of users remain faithful, showing less migration to other currencies.

\subsection{Can we locate individuals associated with the crash?}
The global analysis identified anomalous selling activity on two specific days, namely the 3rd April 2022 and the 19th April 2022. What remains uncertain is whether this activity stems from a widespread increase in overall activity or is driven by a few specific user accounts. Additionally, we aim to ascertain if there is overlap in users across different days, particularly whether the same users are involved in the anomalous days. To investigate, we analysed the token amounts sold on these selected days, aiming to determine each user's contribution to the total sales for a given day. We computed the percentage of total sales attributed to each user, visualising the top 10 users and the remaining are condensed into a single entity, the tail. This visual representation aims to distinguish between scenarios. If the surge in selling activity is due to a general increase in overall activity, we expect to observe a more evenly distributed subdivision among the top 10 users, with a significant portion of sales originating from the remaining users in the tail. Conversely, if the increase in token movements is driven by a few users, we anticipate a diminished impact from users outside the top 10. To validate the anomaly, we conducted a parallel analysis on control days, specifically the start and end of the pre-crash period (2nd April 2022, 1st May 2022), and a day between the two peaks (11th April 2022). In the analysis, we excluded the Zero address (0x0...0) whose outward movements represent token creation (minting). 

\begin{figure}[ht!] 
    \centerline{\includegraphics[width=0.6\textwidth]{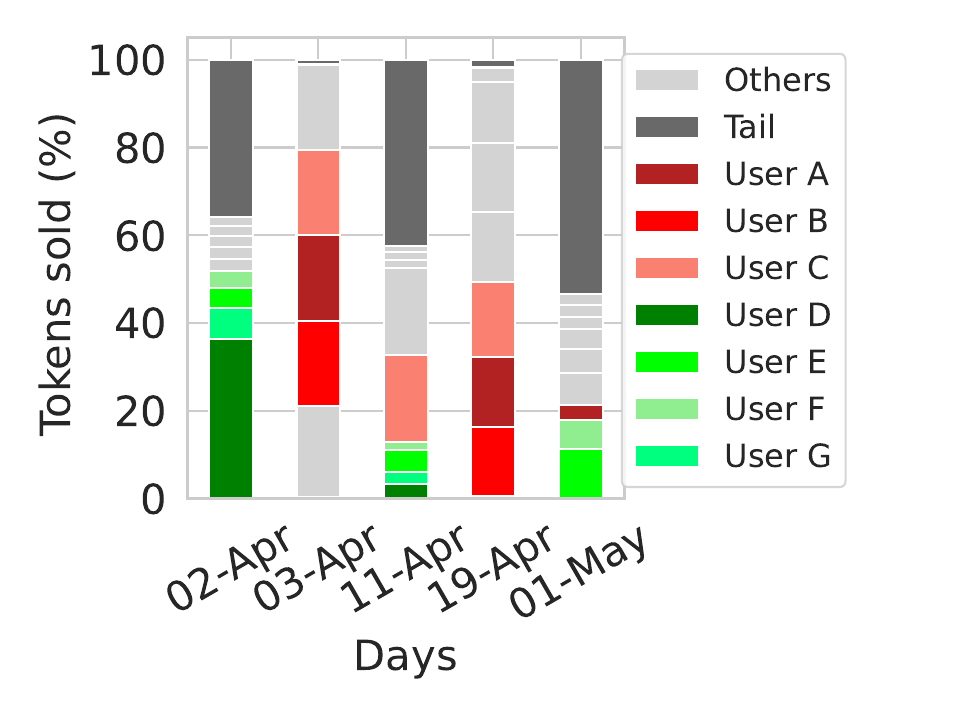}}
\caption{\textit{Top 10 addresses in terms of USTC tokens sold during selected days.} We represent the 2 anomalous days in terms of pre-crash USTC selling, 3rd April 2022 and 19th April 2022, and some control days without anomalies. We represent the percentage of tokens sold over the selected day by the top 10 users, the rest are condensed as the tail. Users from the top 10 that were active on both anomalous days are coloured in red, while users recurrent on other days are coloured in green.}
\label{fig:USTCPeaks}
\end{figure}

\figurename~\ref{fig:USTCPeaks} illustrates the segmentation of the top token sellers on both the anomalous days and the control days. Observations indicate that on peak days (3rd April 2022, 19th April 2022), a few top users dominate the token movement, accounting for a significant percentage of total token sales --- almost $99\%$ on both days. In contrast, on control days, sales are more evenly distributed, with the top 10 users contributing a lesser percentage to total sales, we observe a maximum of around $65\%$. Another important aspect is the presence of recurrent users across the various days. In general, we observe that some users are active across different days. Users A, B, and C consistently appear in the top 10 during the two anomalous days, driving the anomalous peaks. However, their activities on non-peak days are comparatively subdued. Notably, user C displays activity between peaks but not afterwards, resurfacing only on the last control day. Other accounts intermittently appear in the top 10 on regular days, yet their influence is eclipsed by the substantial movements of users A, B, and C. This reveals that among the users contributing to anomalous peaks, some exhibit coordinated efforts, appearing active exclusively on peak days or displaying interconnected movements. Given the extremely unusual distribution of sales (five/six) wallets with approximately the same number of sales dominating $99\%$ of the market there can be little doubt that the individuals (or single individuals or organisation) controlling these wallets is deliberately participating in an attempt to destabilise Luna/Terra. It is also important to remember though that this study is only on those transactions over ethereum and this is discussed in more detail in the next section. 

\subsection{Discussion}

It is important to note the key limitations of this study. The most important one is that we do not directly have access to data from the Luna/Terra ecosystem. The Luna/Terra transactions we study are ``wrapped" transactions in WLUNC/USTC traded on the ethereum system and the original data is not available to researchers. While it is impossible to be one hundred percent certain that the answers to our research questions would transfer directly it seems likely that the majority would. There is no economic reason why the trading patterns should be particularly different. With access to the original data then other wallets would presumably be found associated with the attack as discussed in our final research question. It would be very difficult to prove if the wallets we found belong to one individual, one organisation or multiple individuals and it would be even more difficult to do that across different chains even if that data were available.

How would the techniques used here transfer to investigation of other cryptocurrency attacks? It is very hard to say as each attack has a different nature. The attack investigated here was interesting to study precisely because it was attack based upon selling patterns and attacking the algorithmic behaviour of the cryptocurrency associated with the stablecoin. Other cryptocurrency failures have been based on problems which would certainly manifest differently (for example theft from wallets or security problems). We would not usually expect to be able to identify events such as S1 or S2 leading up to such events. However, the broad range of techniques used in this paper will surely prove useful: correlations between currencies, unusual patterns in trading, mapping the transition between currencies as users lose confidence. The ability to do network analysis at different time-scales, across time and between different layers is really the heart of the approach we advocate rather than the exact analyses included in this paper. 

\section{Conclusions}
\label{sec:conclusions}

This work underscores how novel temporal network analysis can reveal new details in data from crytocurrency transactions, one of the novel paradigms for the Web. It emphasises the importance of separating transactions in layers in the analysis of temporal heterogeneous data, such as transaction data from the web. Using the Raphtory software we were able to simultaneously look at multiple layers (currencies), and multiple timescales and track events as time evolved.

In this study, our focus was on examining the events leading up to and immediately following the collapse of the Luna Terra ecosystem, employing temporal multilayer analysis. This approach enabled a layer-by-layer exploration of the ecosystem, allowing us to assess the crash's impact across currencies. Our research addressed various questions related to the crash, revealing the synchronisation of stablecoin ecosystems before the crash and the new equilibrium found afterwards. We found that the vast majority of users concentrate their activity on just a single layer of the network.  Post-crash we find a small, temporary rally in trading but actually very little change in the majority of traditional graph measures considered. We show that the majority of users trading the affected currencies simply exited trading any of the currencies examined. Finally, we show definite evidence of collusion among a very small number of users in coordinating events leading up to the crash.

\begin{acks}
This work is partly funded by the Cisco research gifts scheme. 
\end{acks}

\bibliographystyle{ACM-Reference-Format}
\bibliography{luna_terra}

\end{document}